\documentclass[a4paper,fleqn,usenatbib]{mnras}
\usepackage[T1]{fontenc}
\usepackage{ae,aecompl}

\usepackage{graphicx}	
\usepackage{amssymb}	

\usepackage{url}
\usepackage{color}
\usepackage{url}
\usepackage{journals}

\newcommand{\msun}{{\rm M}_{\odot}}

\newcommand {\be}{\begin {equation}}
\newcommand {\ee}{\end {equation}}

\newcommand{\src}{1RXS~J180408.9$-$342058}

\title[Jet quenching in \src]
      {Jet quenching in the neutron star low-mass X-ray binary \src}
 \author[Gusinskaia et al.]
    {N.V. Gusinskaia,$^{1}$\thanks{E-mail: N.Gusinskaia@uva.nl} 
          A.T. Deller,$^{2,5}$
          J.W.T. Hessels,$^{1,2}$ 
          N. Degenaar,$^{1,3}$
\newauthor J.C.A. Miller-Jones,$^{4}$  
          R. Wijnands,$^{1}$
          A.S. Parikh,$^{1}$
	  T.D. Russell,$^{1}$
	  D. Altamirano$^{6}$
  \\
$^1$Anton Pannekoek Institute for Astronomy, University of Amsterdam, Science Park 904, 1098 XH Amsterdam, The Netherlands\\
$^2$ASTRON, the Netherlands Institute for Radio Astronomy, Postbus 2, 7990 AA, Dwingeloo, The Netherlands\\
$^3$Institute of Astronomy, University of Cambridge, Madingley Road, Cambridge, CB3 OHA, UK\\
$^4$International Centre for Radio Astronomy Research, Curtin University. GPO Box U1987, Perth, WA 6845, Australia\\
$^5$Centre for Astrophysics and Supercomputing, Swinburne University of Technology, P.O. Box 218, Hawthorn, VIC 3122, Australia\\
$^6$Department of Physics and Astronomy, University of Southampton, Southampton, Hampshire SO17 1BJ, UK}

\date{Accepted xxxx xxx xx.  Received xxx xxxx xx; in original form 2017 April 12}

\pubyear{2017}

\begin{document}
\label{firstpage}
\pagerange{\pageref{firstpage}--\pageref{lastpage}}
\maketitle

\begin{abstract}

We present quasi-simultaneous radio (VLA) and X-ray ({\it Swift})
observations of the neutron star low-mass X-ray binary (NS-LMXB) \src\,
(J1804) during its 2015 outburst.  We found that the radio jet of J1804
was bright ($232 \pm 4$\,$\mu$Jy at 10\,GHz) during the initial hard X-ray
state, before being quenched by more than an order of magnitude during the
soft X-ray state ($19 \pm 4$\,$\mu$Jy). The source then was undetected
in radio (< 13 $\mu$Jy) as it faded to quiescence. In NS-LMXBs,
possible jet quenching has been observed in only three sources and the
J1804 jet quenching we show here is the deepest and clearest example to
date.  Radio observations when the source was fading towards quiescence
($L_X = 10^{34-35}$\,erg\,s$^{-1}$) show that J1804 must follow a steep
track in the radio/X-ray luminosity plane with $\beta > 0.7$ (where
$L_R \propto L_X^{\beta}$). Few other sources have been 
studied in this faint regime, but a steep track is inconsistent with
the suggested behaviour for the recently identified class of transitional
millisecond pulsars. J1804 also shows fainter radio emission 
at $L_X < 10^{35}$\,erg\,s$^{-1}$
than what is typically observed for accreting millisecond pulsars.
This suggests that J1804 is likely not an accreting X-ray or transitional
millisecond pulsar.

\end{abstract}

\begin{keywords}

{Low mass X-ray binary -- Radio/X-ray correlation -- Disk/jet coupling
  -- Sources, individual: \src}

\end{keywords}

\section{Introduction}
\label{sec:intro}


Accretion is a fundamental astrophysical process that occurs in a very
broad range of astrophysical contexts, e.g. from proto-stars and
planetary systems, to stellar-mass compact objects in short-orbit
binaries, to active galactic nuclei (AGNe).  Low-mass X-ray binaries
(LMXBs) are binary systems with a neutron star (NS) or stellar-mass black
hole (BH) primary and a low-mass ($< 1\, \msun$) secondary star (which may
be semi-degenerate or not).  Most of the time, LMXBs are in a
so-called quiescent state (where the X-ray luminosity, $L_X$, 
is $<\, 10^{34}$\,erg/s). Many of these systems undergo episodic
outbursts, where the accretion rate increases
by orders of magnitude, and the system exhibits enhanced
radiation. During outburst, emission is typically the strongest in the X-ray band (usually
associated with the accretion inflow, disk and corona; \citealt{SS1973}, \citealt{Begelman1983}) and 
radio emission is often observed, indicative of a jet or another type of outflow \citep{FENBELGAL2004}.
LMXB outbursts have time scales of weeks to years, allowing us to investigate
accretion in detail and in a dynamic way. LMXBs with black hole primaries (BH-LMXBs)
are radio brighter at a given $L_X$ and are to date arguably better characterized in the radio 
in comparison to NS-LMXBs (\citealt{FENKUUL2001}; \citealt{MIGFEN2006}).

A typical BH-LMXB outburst starts in a hard X-ray spectral state,
where the X-ray spectrum is dominated by a hard power law component
(see \citealt{McC2004} and references therein). In some models, this
state is characterised by a truncated accretion disk, where soft X-ray
photons from the disk are Comptonized in an electron corona around the
compact object (see e.g. \citealt{Esin1997}). In this state, radio emission
is often detected and is associated with optically thick synchrotron radiation
from a steady, self-absorbed, compact jet (e.g. \citealt{Fen2004}, \citealt{Fen2009}).
As the outburst progresses, the disk brightens due to an increased mass accretion rate.
As the X-ray luminosity increases, the inner accretion disk moves inwards, closer to the
central object (see e.g. \citealt{Malzac2007}). The X-ray emission becomes dominated by
a soft thermal component (described by blackbody emission from
the hot inner region of the accretion disk), as the system moves towards the soft X-ray state. During
this transition the brightest radio emission is observed. This emission is associated with optically-thin
radio ejecta moving away from the system, which have been spatially resolved in a few BH-LMXBs
(e.g. \citealt{Mirabel1994}). At this time, the steady jet is quenched (\citealt{Russell2011};
\citealt{Corbel2004}), appearing again in the reverse transition back to the hard state.
At the peak of the outburst, BH-LMXBs can make several transitions between the hard and soft
states \citep{Fen2004}.

NS-LMXBs often exhibit hard and soft X-ray states similar those seen
in BH-LMXBs.  Historically, there are two classes of NS-LMXBs, defined by their
shape in an X-ray color-color diagram: Z-type and Atoll-type NS-LMXBs
\citep{HV1989}.  Z-type sources accrete at higher accretion rates and generally
have softer spectra than Atoll-type NSs. However, some transient sources have
exhibited both Atoll and Z-type behaviour at low and high accretion rates, respectively
(e.g. \citealt{Homan2010}; \citealt{Homan2014}).  This indicates that these two types
of behaviour are driven by a changing mass accretion rate and not a fundamental
difference in source type.

Two other subclasses of NS-LMXBs are accreting millisecond X-ray pulsars (AMXPs) and
transitional millisecond pulsars (tMSPs). During their outburst phases, AMXPs exhibit
coherent X-ray pulsations as a result of the accretion flow being channeled by the
NS's magnetic field onto its polar caps \citep{Patruno2012}.
tMSPs are NSs that switch between rotation-powered radio pulsar
and accretion-powered LMXB phases (currently, only three confirmed transitional sources
are known; see \citealt{Bogdanov2015}). Similar to Atoll and Z-type NSs, AMXPs and tMSPs
could also be different manifestations of the same source class, as implied by M28I
exhibiting both AMXP and tMSP behaviour at different mass accretion rates \citep{Papitto2013}. 

Simultaneously observing LMXBs in the X-ray and radio bands allows us to explore connections
between the inflow and outflow of material, investigating the relationship between radio brightness and
the X-ray spectral state. Comparing the radio/X-ray correlation for different classes
of accreting binaries allows us to investigate the influence of physical characteristics such
as mass of the accretor, absence or presence of a stellar surface, and the strength of
the magnetic field, on jet launching.

\begin{figure*}
\centering 
\includegraphics[width=18.5cm]{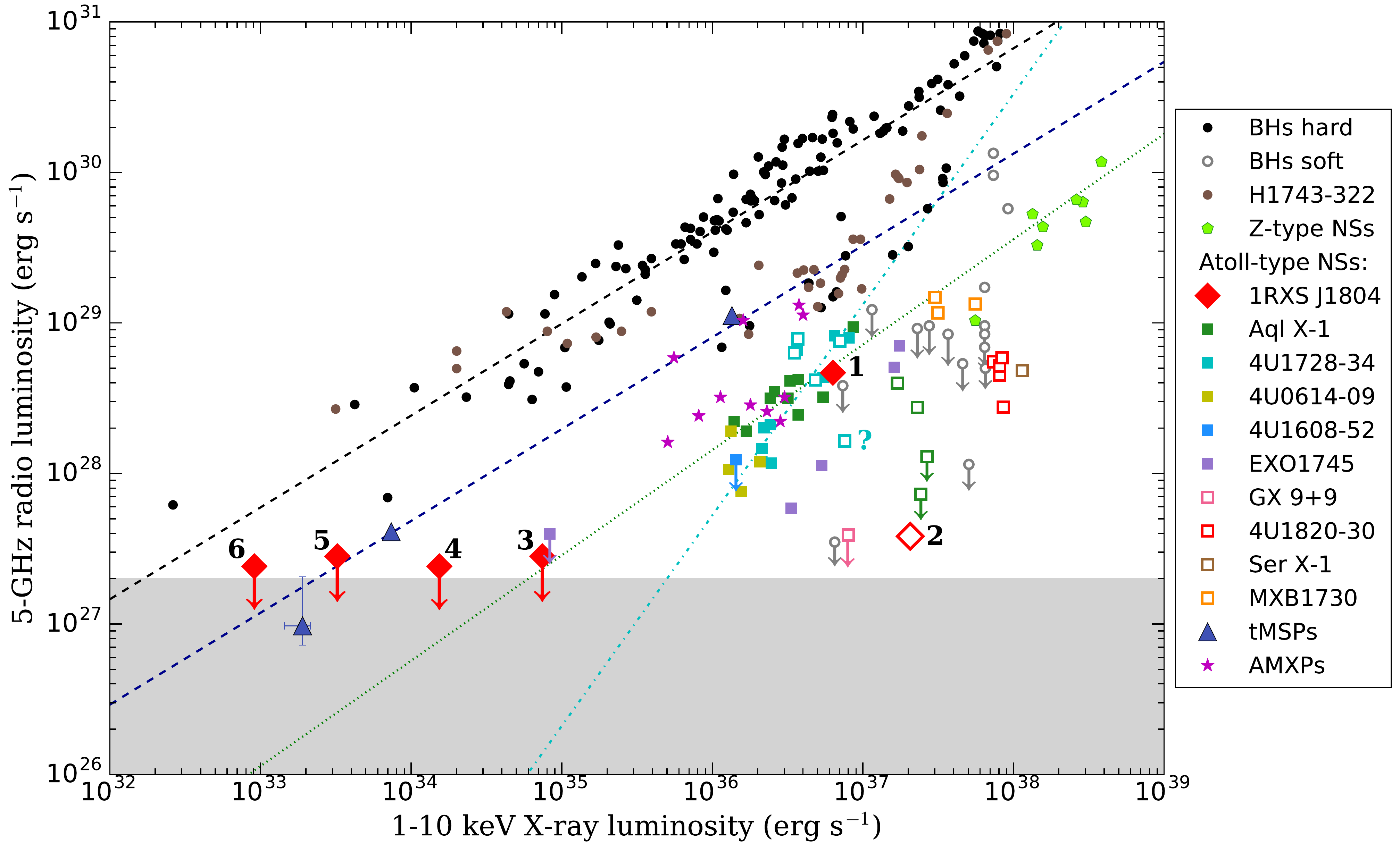} 

\caption{X-ray ($1-10$\,keV) luminosity versus radio (5\,GHz)
  luminosity for BH- and NS-LMXBs.  The grey shaded area indicates the
  region of the parameter space that is effectively inaccessible given
  the sensitivity of current radio telescopes (15 $\mu$Jy for $\sim$ 1 h integration time) for sources at a typical
  distance of $\sim 5$ kpc (note that tMSP PSR~J1023+0038, the lowest triangle, is
  exceptionally close at only 1.3\,kpc, which is why it is the only system
  detected in this region). Circles represent BH-LMXBs; pentagons represent Z-type
  NS-LMXBs, squares represent Atoll-type NS-LMXBs, and for these we give individual names.
  Triangles represent tMSPs and stars represent AMXPs. J1804 is represented by red diamonds.
  Filled symbols represent hard X-ray state observations; open symbols
  represent soft state X-ray observations.  For the single cyan point
  marked with `?', it is unclear whether this observation was truly in
  the soft state. The black dashed line represents the radiatively inefficient $L_R \propto L_X^{\beta}$
  track of BH-LMXBs, with $\beta \approx 0.6$. The dark blue dashed line represents the tentative
  $\beta\approx$0.6 tMSP track proposed by \citet{DEL2015}. The dark green dotted line
  represents the $\beta\approx$0.7 track derived for Aql~X-1 by \citet{TUD2009} and \citet{MJ2010}.
  The cyan dash-dotted line represents the $\beta\approx$1.4 track derived for 4U1728$-$34 by
  \citet{MIG2003}. Data points are taken from: \citet{Coriat2012} for BHs in the hard X-ray state;
  \citet{Coriat2011} for H1743$-$322 and most of the BHs in the soft X-ray state;
  \citet{Gierlinski2001} and \citet{Russell2011} for the two lowest upper limits of BHs in the soft state;
  \citet{MIGFEN2006} for Z-type NSs; \citet{TUD2009} and \citet{MJ2010} for Aql~X-1;
  \citet{MIG2003} for 4U1728$-$34; \citet{MIG2004} for 4U1820$-$30 and Ser~X-1;
  \citet{Moore2000} and \citet{Kuulkers2003} for MXB~1730$-$335;
  \citet{MIG2011} for GX~9+9; \citet{Tetarenko2016} for EXO~1745$-$248;
  \citet{DEL2015}, \citet{Hill2011} and \citet{Papitto2013} for tMSPs;
  \citet{MIG2005} for 4U1608$-$52, 4U0614$-$09 and the AMXPs.}\label{fig:1}

\end{figure*}

For BH-LMXBs there is a correlation between the radio ($L_R$) and X-ray ($L_X$)
luminosities in the hard X-ray state, when the radio jet is steady, all the way
down to quiescence (\citealt{Corbel2000}; \citealt{GALLO2003}; \citealt{Gallo2006}; \citealt{Gallo2014}).
This relation is described by a powerlaw, such that $L_R \propto L_X ^{\beta}$, where $\beta \sim 0.63$,
which holds over more than eight orders of magnitude in X-ray luminosity and $\sim$four-five
orders of magnitude in radio luminosity (see \citealt{GALLO2003} and
black circles in Fig. \ref{fig:1}). This correlation can be expanded
to encompass supermassive BHs and AGNe by using a mass scaling (the
so-called `fundamental plane of BH activity'; for more details see
\citealt{Merloni2003}; \citealt{Falcke2004}; \citealt{Plotkin2012}).
This power-law index is thought to be due to either accretion onto a BH being
radiatively inefficient because infalling matter and energy is able to pass
beyond the event horizon \citep{Narayan1995}, or because BHs channel a significant fraction of
the accretion energy into the jets \citep{FenGalJon2003}. However, there are also
`radio-quiet outliers' (e.g. H1743$-$322, Figure~\ref{fig:1}; \citealt{Coriat2011})
which show a steeper correlation, where $\beta \approx 1-1.4$ at higher X-ray
luminosities ($L_X \geq 10^{36}$ erg/s; \citealt{Gallo2012}), implying
radiatively-efficient accretion (see Figure~\ref{fig:1}). At lower X-ray luminosities these objects
appear to rejoin the `standard' radiatively-inefficient BH track (\citealt{Maccarone2012};
\citealt{MeyerHof2014}).

In general, NS-LMXBs show weaker radio jets \citep{MIGFEN2006} than BH-LMXBs, with flux
densities ranging from hundreds of $\mu$Jy (for typical distances of $\sim
5$\,kpc) down to only tens of $\mu$Jy, pushing the observational
capabilities of even the most sensitive radio
interferometers. Consequently, the connection between X-ray and radio
emission in NS-LMXBs remains relatively poorly studied compared to
BH-LMXBs (\citealt{MIGFEN2006}, \citealt{Fen2016}).

For NS-LMXBs the radio/X-ray luminosity correlation has been well
monitored for only three sources so far, showing a wide range of powerlaw
indices (4U1728$-$34: $\beta=1.4 \pm 0.2$ \citealt{MIG2003}; Aql~X-1: $\beta=0.76 \pm 0.1$
and EXO 1745$-$248: $\beta=1.7 \pm 0.1$; \citealt{Tetarenko2016}). However,
caution should be taken when interpreting these differences because in some
cases the fits are based on a narrow luminosity range ($\sim 1$ dex). For the complete
sample of all Atoll-type NS-LMXBs (those which have been observed
simultaneously in the X-ray and radio), an overall correlation was found,
showing an index of $\beta \sim 1.4$ \citep{MIGFEN2006}, however again the luminosity
range studied is also limited. This steep slope implies that NS-LMXB accretion is
radiatively efficient (see cyan dashed line in Fig.\ref{fig:1}), consistent
with the idea that accreting material falling onto the surface of the neutron
star naturally produces radiation. In fact, NS-LMXBs in the hard state appear
to follow a similar radiatively efficient track to the `outlier' BH-LMXBs, in
which the origin of the radiatively-efficient accretion is still not well-established.


In addition to the $L_R/L_X$ hard state correlation, BH-LMXBs show
radio flaring before, or at the time of, the transition from the hard to the soft state,
preceded by the quenching of the compact jet (\citealt{Fen1999}, \citealt{Fen2009},
\citealt{GALLO2003}, \citealt{MJ2012}). For most BH-LMXBs
no radio emission has been observed in the soft X-ray state. However, in a few
sources there have been some detections of faint radio emission with an inverted
spectrum, most likely originating from optically-thin residual jet emission \citep{Fen2009}.
At a similar X-ray luminosity, hard and soft-state radio emission can differ in brightness
by approximately three orders of magnitude (e.g. by $> 700\times$ for H1743$-$322; \citealt{Coriat2011}).


In NS-LMXBs the jet quenching phenomenon is the topic of considerable
debate (\citealt{Fen2016}; \citealt{MIG2011}). To date, only two NS-LMXBs
have been observed simultaneously in the X-ray and radio bands during both
hard and soft X-ray states: 4U1728$-$34 \citep{MIG2003} and Aql~X-1 (\citealt{TUD2009},
\citealt{MJ2010}). For 4U1728$-$34, only marginal evidence for jet quenching was
observed, while in Aql~X-1 the jet was quenched by at least one order of magnitude.
Furthermore, there are four systems which were observed at radio frequencies only while
in the soft state: 4U1820$-$30, Ser~X-1; \citep{MIG2004}, MXB~1730$-$355; (\citealt{Rutledge1998},
\citealt{Kuulkers2003}) and GX~9+9\, \citep{MIG2011}. 4U1820$-$30, Ser~X-1 and  MXB~1730$-$355 showed
surprisingly strong radio emission (if the assumed distance is accurate), albeit
at a slightly lower radio luminosity than other hard-state NS-LMXBs (see Fig.\ref{fig:1}),
while no radio
emission was detected for GX~9+9\, \citep{MIG2011}. It is important to note that in the case
of Atoll-type NS-LMXBs there have been many detections in the soft state (6 out of the
9 systems observed in the radio band), while in BH-LMXBs almost all radio observations
in the soft state have provided non-detections. Bearing in mind that NS-LMXBs
tend to be less radio loud than BH-LMXBs, the observed level of jet quenching in
NS-LMXBs seems to be less extreme than in BH-LMXBs \citep{MIGFEN2006}.

\subsection{\src}


\src\ (hereafter J1804) was classified as a NS-LMXB system on 16 April
2012 when {\it INTEGRAL} detected a Type I X-ray burst from a
previously unclassified X-ray source \citep{ATel_INTEGRAL2012}.
Assuming that the Eddington luminosity was reached during the burst \citep{Kuulkers2003},
the upper-limit to the distance was estimated to be $\sim 5.8$\,kpc \citep{ATel_INTEGRAL2012}.
{\it Swift}/XRT observations on 17 and 30 April 2012 revealed faint X-ray
emission (${L_{X, {\rm  0.5-10\, keV}} \sim 10^{33-34}}$\,erg/s),
consistent with a very low rate ($\sim 10^{-4} L_{{\rm Edd}}$) of accretion
\citep{ATel_Quiescence2012}. Based on this, the source was classified as
a very faint X-ray transient (VFXT; see \citealt{Wijnands2006} for
classification details).


On 27 January 2015 the {\it Swift}/BAT hard X-ray monitor
($15-50$\,keV; \citealt{Matsuoka2009}) detected a new, bright outburst
from J1804 \citep{ATel_Swift2015_detecion}, which was later confirmed by {\it MAXI/GSC}
($2-20$\,keV;\, \citealt{ATel_MAXI2015}).  Subsequent {\it Swift}/BAT observations
revealed steady, hard X-ray emission, consistent with canonical
accreting LMXB outbursts (${L_{X, {\rm 15-50 keV}} \sim
  10^{36-38}}$\,erg/s).  During this outburst, J1804 was monitored in a
multiwavelength campaign (X-ray: \citealt{INTEGRAL2015_ATel},
\citealt{Ludlam2016},\citealt{Parikh2016}; optical and IR: \citealt{BAG2016}).
In April 2015, J1804 transitioned to the
soft X-ray state \citep{ATel_DEG2015} before returning to quiescence at
the beginning of June 2015 \citep{Parikh2016}.


It has been suggested (e.g. \citealt{Deg2014} and \citealt{Heinke2015})
that some VFXTs may be similar in nature to tMSPs. Furthermore, the X-ray
spectrum of J1804 in its initial hard state is consistent with a $\Gamma \sim 1$
power law \citep{ATel_Swift2015}, which is strikingly hard for typical NS-LMXBs
\citep{Parikh2017} but similar to that displayed by the tMSPs
\citep{Tendulkar2014}. The tMSPs have also been found to be surprisingly
bright in radio, even at low X-ray luminosities \citep{DEL2015}.

In order to investigate the X-ray/radio behaviour of J1804, we performed simultaneous observations with {\it Swift}/XRT and the Karl G. Jansky
Very Large Array (VLA) radio interferometer, monitoring the source
through its 2015 outburst and transition back to quiescence.  Our aim was
to track J1804 in the $L_R/L_X$ diagram, comparing it to typical hard state NS-LMXBs,
as well as tMSPs and AMXPs.


\cite{BAG2016} investigated the near-infrared(NIR)/optical/ultraviolet (UV) spectrum of J1804
during its 2015 outburst. On 26 February 2015, during the hard X-ray state, they found a
low-frequency excess in the NIR band, indicating the presence of a jet. Their observations on
24 April 2015, when J1804 was in the soft X-ray state, showed that the NIR excess had disappeared,
thus indicating a reduction of the jet emission.


\cite{Ludlam2016} observed J1804 with the {\it NuSTAR} and {\it
  XMM-Newton} X-ray satellites during its 2015 hard X-ray state, finding
clear evidence for an Fe $K_{\alpha}$ line and N VII, O VII and O VIII
reflection lines. Using a relativistic reflection model they estimated
the inclination of the binary to be $i\, \sim 18^{\circ}-29^{\circ}$ and
the inner radius of the disk $R_{in} \le 22$km. They found no evidence for
coherent X-ray pulsations, indicating that J1804 is apparently a non-pulsating NS-LMXB.
Using {\it NuSTAR} and {\it Chandra} observations, \citet{Deg2016} detected the Fe $K_{\alpha}$ line
during the soft state, finding similar values for the inclination and inner
radius of the disk: $i \sim27^{\circ} - 35^{\circ}$, $R_{in} \le 11 - 17$\, km.

In this paper we present the data reduction (\S\ref{sec:obs}) and the results of the
combined radio/X-ray analysis of J1804 (\S\ref{sec:results}), discussing
its implications in comparison to other NS-LMXB systems (\S\ref{sec:disc}).

\section{Observations and data analysis}

\label{sec:obs}


We performed quasi-simultaneous observations of J1804, using the {\it Swift}
X-ray telescope and the VLA radio interferometer, during its
outburst in March -- June 2015.  Table~\ref{obs_all} provides a
log of all observations used in this work.


\begin{table*}

\caption[All observations]{The VLA 10-GHz radio detections and upper
  limits (3 $\sigma$) together with the corresponding
  quasi-simultaneous {\it Swift} X-ray observations and
  spectrum.} \label{obs_all}

\begin{minipage}{170mm}
\begin{tabular}{lccccccccc}
\hline
\hline
Date & MJD & Obs. ID   & Unabs. Flux $^a$  &  Photon &      &       & Date   & MJD & Flux density $^b$   \\
Swift & Swift  & and       & X-ray &  index  & $\chi^2_{\nu}$ & dof & VLA    & VLA &    radio       \\     
(2015)&   (d)  & obs. mode & ($10^{-9} \mathrm{erg/cm^2/s}$)& & & & (2015) & (d)  & ($\mu$Jy) \\ 
\hline
16 Mar& 57097.29  &  00032436024(WT) & $1.618 \pm 0.026$ & $1.12 \pm 0.02$ & 1.01 & 523 &17 Mar& 57098.59  & 232 $\pm$ 4\\
19 Mar& 57100.35  &  00032436025(WT) & $1.507 \pm 0.025$ & $1.18 \pm 0.02$ & 1.08 & 503 &    -      &       -     \\
12 Apr& 57124.17  &  00032436032(WT) & $4.959^{+0.076}_{-0.075}$ & $1.62 \pm 0.04$ & 1.06$^c$ & 502 &13 Apr& 57125.51  &  19 $\pm$ 4 \\
14 Apr& 57126.55  &  00081451001(WT) & $5.252^{+0.052}_{-0.051}$ & $1.66 \pm 0.03$ & 1.01$^c$ & 599 &   -    &       -     \\
1 Jun & 57174.65  &  00032436035(PC) & $4.120^{+0.45}_{-0.43} \times 10^{-2}$ & $1.95 \pm 0.15$ & 1.02 & 42 &1 Jun & 57175.39  &       < 14  \\
2 Jun & 57175.58  &  00033806001(PC) & $1.502^{+0.15}_{-0.14} \times 10^{-2}$ & $1.99 \pm 0.12$ & 1.09 & 43 &2 Jun& 57177.35  &       < 12  \\
7 Jun & 57180.26  &  00033806002(PC) & $4.01^{+2.81}_{-2.14} \times 10^{-4}$ & $3.17 \pm 1.01$ & 0.53 & 16 &7 Jun& 57179.37  &       < 14  \\
9 Jun & 57182.63  &  00032436037(PC) & $2.12^{+7.67}_{-1.66} \times 10^{-4}$ & $3.67^{+2.09}_{-2.79}$ & 0.29 & 16 &   -      &       -     \\
11 Jun& 57184.58  &  00033806003(PC) & $2.25^{+1.75}_{-1.63} \times 10^{-4}$ & $2.56^{+0.99}_{-1.15}$ & 0.99 & 13 & 11 Jun& 57184.27  &       < 12   \\
\hline
\hline
\end{tabular}
\begin{flushleft}{
$^{a}$ X-ray flux in 1--10 keV range (90\% confidence interval)\\
$^{b}$ Radio flux density at 10 GHz with 1 $\sigma$ errors.\\
$^{c}$ For 12 Apr and 14 Apr observations we used an additional blackbody spectral component with temperatures of 0.89$\pm0.07$ keV and 0.89$\pm$0.04 keV respectively.
  }\end{flushleft} 
\end{minipage}
\label{tab:obs}
\end{table*}


\subsection{Radio Data (VLA)}


J1804 was observed with the VLA at 6 epochs (project ID: 15A-455): 
each at X-band (8--12 GHz) for a total duration of $\sim 1$\,h including calibration scans,
with $\sim 30$\,min on source.  During the first 2 epochs in March and
April 2015 the array was in B configuration (synthesized beam $\sim
1.25 $\,\arcsec), and during the last four epochs the array was
transitioning to BnA, gradually increasing in resolution up to $\sim
0.6$\,\arcsec. We used J1331+305 and J1407+2807 as flux and
polarization calibrators, and we used J1806$-$3722 as a phase and
amplitude calibrator.


The data were reduced using CASA \citep{CASA2007}  and the
EVLA pipeline. We combined all epochs in one measurement set
and used the multi-scale/multi-frequency clean method in CASA
\citep{MSCLEAN2008} to make a source model of the field.  We then used
this model to apply self-calibration corrections to each epoch before
imaging and extracting the target flux density (or upper limit).  We
used phase self-calibration with a solution interval of 12 seconds,
and amplitude self-calibration with an interval of 10 minutes.  The
RMS noise of the per-epoch images was $3-4$\,$\mu$Jy, and the combined
image of all epochs reached a root mean square (RMS) noise of $\sim 2$\,$\mu$Jy.

For the epochs where the source was detected, we performed spectral and
time-series analyses. In the first radio observation we split the 8--12 GHz band into 8 spectral bins, each consisting of 4 subbands, where a subband is a 128-MHz chunk
of data comprised of 64 x 2-MHz channels. We imaged each subband with the source model
obtained from our full-band imaging at this epoch. For our second observation we applied
the same technique; however, due to the faintness of the radio source, we split the band
into two 2-GHz subbands.

\begin{figure*}
\centering 
\includegraphics[width=15cm]{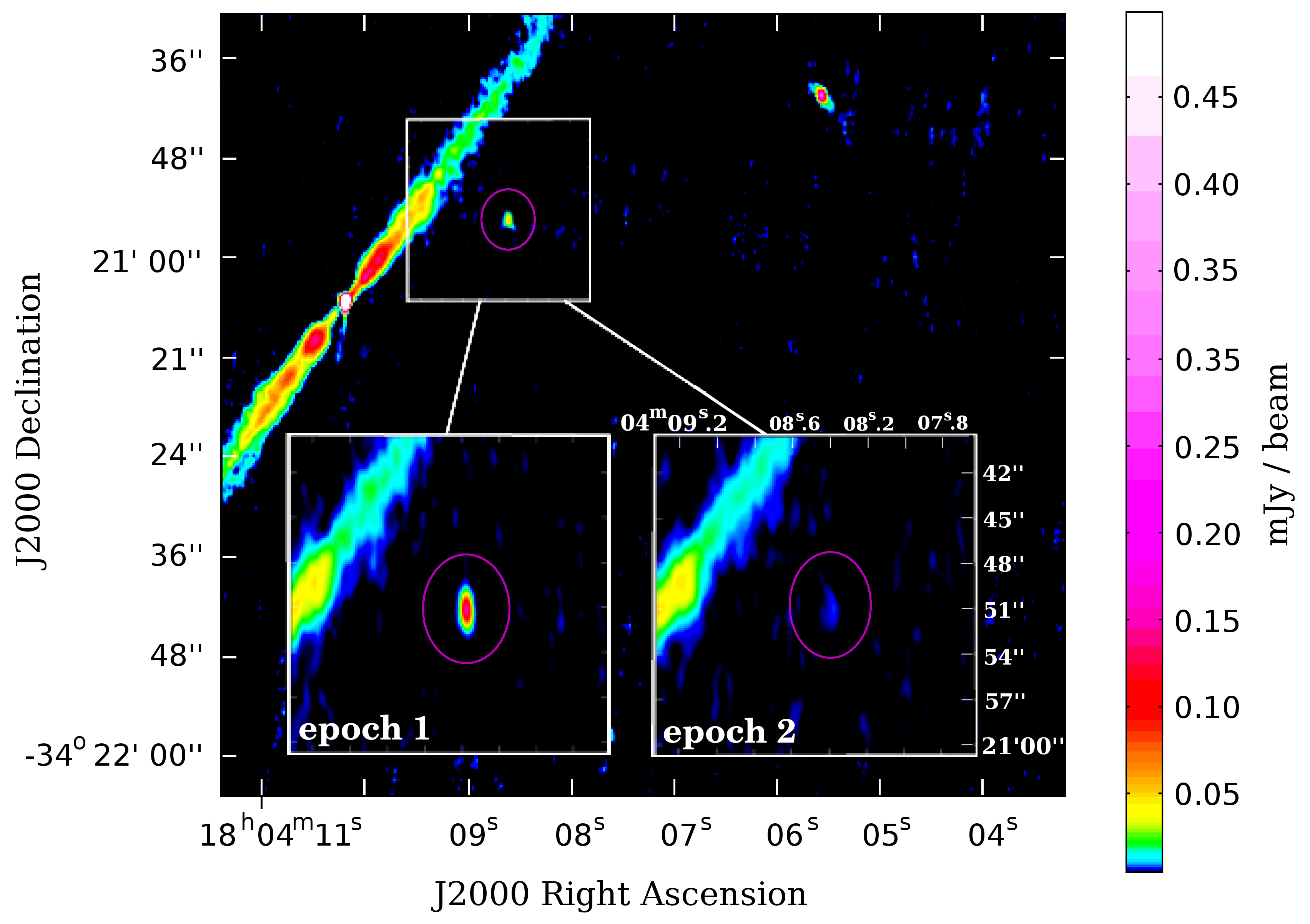} 

\caption{VLA combined radio image (all 6 epochs) of J1804 and the
  surrounding field.  The extended diagonal feature corresponds to the lobes of a
  nearby radio galaxy (NVSS~J180414$-$342238).  The insets at the bottom show images from the
  first (left) and second (right), epochs individually.}\label{fig:2}
\end{figure*}

Figure \ref{fig:2} shows the combined (8 -- 12 GHz) radio interferometric map of J1804
and its surroundings, where all 6 observation epochs have been stacked together.
In all VLA observations J1804 was unresolved within the $1.9 \times 0.65$ arcsecond beam.
J1804 is located close to the radio galaxy NVSS~J180414$-$342238 \citep{CON1998}, whose
extended lobes are prominently visible in the upper-left side of the radio image (see
Figure~\ref{fig:2}). The combination of all observations allowed us to create the best
model of the extended emission close to the source, allowing us to achieve a sensitivity
close to the theoretical thermal noise.

\subsection{ X-rays ({\it Swift, MAXI})}


Based on the 6 observation epochs obtained with the VLA, we selected 9
quasi-simultaneous {\it Swift}/XRT observations (Target ID 32436),
i.e. each radio observation is framed by a pair of X-ray observations
(for more details see Table~\ref{tab:obs}, Figure~\ref{fig:3} and Section 3.3).

In the first 4 observations (March $-$ April 2015), J1804 was bright
(unabsorbed $1-10$\,keV flux $> 10^{-9}$\,erg/cm$^2$/s) and {\it
  Swift} operated in Window Timing (WT) mode with an exposure time per observation of
$\sim 1$\,ks. As J1804 returned to quiescence (unabsorbed $1-10$\,keV flux $< 10^{-12}$\,erg/cm$^2$/s),
the source became faint, such that our final 5 observations were done in
Photon Counting (PC) mode with an individual exposure time of $\sim 2$\,ks.
For more details about the X-ray light curve see \citet{Parikh2016}.


We used the XRT$\_$PIPELINE task for basic data reduction and
calibration.  We extracted the spectrum using {\tt xselect}.  We used
variable in size extraction regions for each WT epoch, in order to properly
correct for {\it Swift}'s roll angle in these 1-D data sets. For each WT
data set, the source and background regions were of equal size.  The
size of WT-mode event regions was no greater than 30 pixel.  In the
PC-mode observations we used the same source and background regions
(a 30-pixel radius circle area for the source, and a 30-pixel width
annulus area for the background with a 30-pixels inner radius). We examined
observations that have a count rate more than $150-200$\,counts/s in WT-mode and $> 0.6$\,counts/s
for PC mode in order to check for possible pile-up (following the method in \citealt{ROM2006}),
and excluded the 3 central pixels for the 3rd and 4th observations.

We used the {\tt XRTMKARF} script together with the `swxwt0to2s6$\_$20131212v015.rmf'
response file to produce an ancillary file for each individual observation (using individual
exposure maps), correcting for known artifacts on the CCD. The {\tt xselect} tool was used to
extract the light curve and flux. We selected photons from the $0.7 - 10$\,keV range for WT
mode observations and $0.4 -10$\,keV for PC mode observations in order to perform the best possible spectral fit with {\tt Xspec}. For the first two observations (in March 2015) an
acceptable fit was achieved using an absorbed power law model ({\tt TBabs*powerlaw} in {\tt Xspec}).
For the two observations in April 2015, a combination of absorbed power law and black body
({\tt bbody} model in {\tt Xspec} with $\sim$1 keV temperature) models provided a better fit
($\chi_{\nu}^2 \approx 1.0$, dof$\approx$500) than a single power law model ($\chi_{\nu}^2 \approx 1.8$).
Parameters of the model and reduced $\chi^2$ value for each epoch are summarized in Table~1.
We performed a simultaneous fit for hydrogen column density (nH) value (using {\tt TBabs} model)
at all epochs in order to get the best estimate, which was found to be $0.36 \pm 0.02 \times 10^{22}$\,cm$^{-2}$.
This value is consistent with previous work on this source (e.g. \citealt{Ludlam2016}, \citealt{Deg2016}).
We used {\tt cflux} in {\tt Xspec} to determine unabsorbed $1-10$\,keV fluxes and their
associated errors for each epoch (see Table~1).


In order to produce a hardness-intensity diagram (HID) for the 2015 outburst of J1804
(see Figure~\ref{fig:4}), we used publicly available data from {\it MAXI}
($2-10$\,keV)\footnote{\url{http://maxi.riken.jp/top/index.php?cid=1&jname=J1804-343}}
and {\it Swift}/BAT ($15-50$\,keV)\footnote{\url{http://swift.gsfc.nasa.gov/results/transients/weak/1RXSJ180408.9-342058/}}.
For both bands we converted count rates into Crab flux units assuming a Crab-like spectrum for J1804. We defined
hardness ratio as $\mathrm{Flux_{(15-50keV)}/Flux_{(2-10keV)}}$ and
intensity as $\mathrm{Flux_{(15-50keV)} + Flux_{(2-10keV)}}$.
 
\section{Results} 
\label{sec:results}

\subsection{Radio evolution}

J1804 was detected by the VLA in our first two observations. At
10-GHz we measured flux densities of $0.232 \pm 0.004$\,mJy and $0.019 \pm 0.004$\,mJy
($> 4\sigma$ detection; see insets of Figure~\ref{fig:2}), respectively.
In the remaining four observations, J1804 was undetected at the best-fit
position from observation 1 -- with 3-$\sigma$ flux density upper limits
of $\sim 0.012-0.014$\,mJy. The source was also undetected in a combined image
of the last four observations, or any combination of these
observations.


We investigated the first and second radio observations (when the source
was detected) for variability and spectral shape. In the first epoch no significant variability
on a $\sim$1 hour time scale was observed. We also observed no frequency dependence and
the 8 -- 12 GHz radio spectrum was consistent with being flat,  with a powerlaw index of $\alpha =
0.12 \pm 0.18$ (where $S_{\nu} \propto \nu^{\alpha}$) in the first observation.
During the second radio observation, J1804 was too faint to establish if it was variable
over the 1-hour observation timescale. Also, the radio spectrum was not constrained in a
meaningful way ($\alpha =-0.31 \pm 0.68$), such that we could not distinguish between a flat or 
steep radio spectrum, and we could not identify whether the radio emission was optically-thick or 
optically-thin.

Using the B-configuration VLA beam size ($1.9 \times 0.65$ arcsecond) and assuming
a distance of 5.8\,kpc, we estimate an upper limit of $<7\times 10^3$ AU (3$\times 10^6$
light-seconds) for the projected linear size of the radio source. Assuming the brightness
temperature does not exceed $10^{12}$\,K (see \citealt{MIGFEN2006}), we also estimate a
lower size limit of $> 0.01$ AU (5.2\,light-seconds) for the first epoch and $> 0.003$ AU
(1.48\,light-seconds) for the second epoch. Using a $\sim 3 \mathrm{h}$ orbital period (tentatively
proposed by \citealt{Deg2016}) and assuming a total system mass of 2 $M_{\odot}$
(1.4 $M_{\odot}$ neutron star and 0.6 $M_{\odot}$ companion), we determine the J1804 binary
would have a semi-major axis of 0.006 AU or $\sim 3$ light-seconds.
In this scenario, the radio emitting region during the first epoch would be larger than the
binary separation and the jet would be gravitationally unbound from the system.

Assuming a flat radio spectrum and 5.8-kpc distance we calculate the 5-GHz
luminosity\footnote{$L_R=4 \pi d^2 \nu S_{\nu}$, where d is distance, $\nu$=5 GHz is the
reference frequency and $S_{\nu}$ is the flux density at the reference frequency.} of J1804 and compare it
to a large sample of BH- and NS-LMXBs (Figure~\ref{fig:1}).

\subsection{Source Position}


The radio detection in the first epoch provides the most precise
position of J1804 to date:\\
 
\noindent RA: $18^{{\rm h}}04^{{\rm m}}08^{{\rm s}}.3747 \pm 0^{{\rm s}}.0002 $,\\
Dec: $-34^{\circ}20'51''.18 \pm 0''.01$.\\

Here the fit errors represent the beam size (for
the given VLA configuration) divided by signal to noise ratio. These 
do not account for the uncertainty in the absolute positional accuracy
of the phase-referenced VLA observations. However, the reference source
J1806$-$3722 is accurate to $\sim$ 1 mas (defined at 8\,GHz), and is only
a few degrees from J1804, meaning that any systematic offset in the absolute
positions are at worst comparable to the positional errors derived from the
VLA beam. This position is about three orders of magnitude more precise than
the previous best position provided by optical \citep{ATel_BAG2015}
and ultra-violet \citep{ATel_Swift2015} observations and consistent with
the position provided by \citet{ATel_DEL2015}.

\subsection{\it Swift}

\begin{figure*}
\centering 
\includegraphics[width=14.5cm]{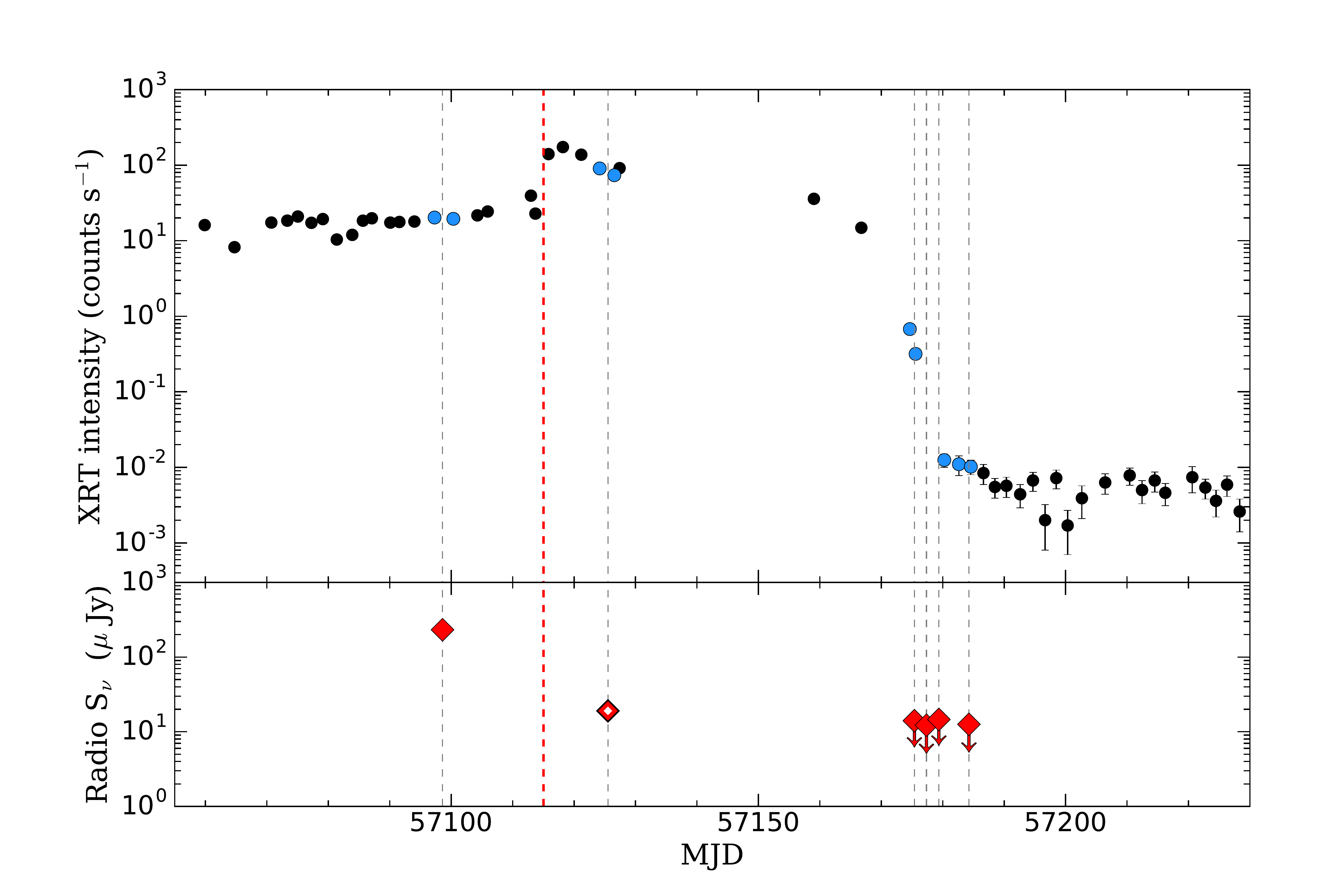} 

\caption{{\it Top}: {\it Swift}/XRT ($0.5-10$\,keV) X-ray lightcurve of the 2015 outburst of J1804
  (data from \citealt{Parikh2016}). {\it Bottom}: VLA (10 GHz) radio lightcurve (S$_{\nu}$ is radio 
  flux density). For each radio observation we selected the two closest X-ray observations
  in time (blue circles) and reanalyzed those data (Table~1). The grey dashed vertical lines correspond to the MJDs
  of the radio observations. The red dashed line corresponds to the approximate time of the hard-to-soft X-ray state transition.}\label{fig:3}

\end{figure*}

\begin{figure*}
\centering 
\includegraphics[width=14.0cm]{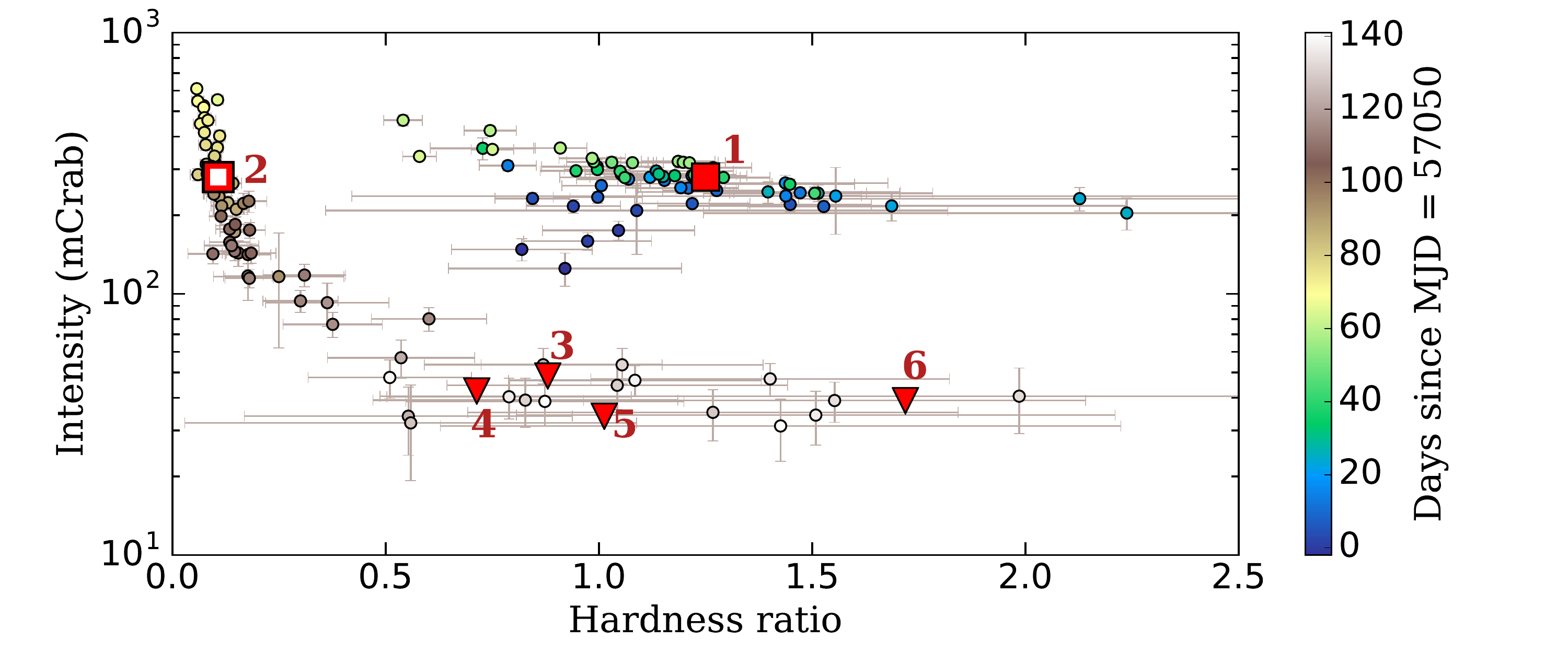} 

\caption{Hardness-intensity diagram of the 2015 outburst of J1804. Red squares represent VLA radio
  detections, downwards triangles represent radio upper limits.  Filled/open symbols indicate radio
  detections in the hard/soft X-ray state. Circles represent X-ray, {\it Swift}/BAT and {\it MAXI} observations, with
  1 $\sigma$ errorbars. The colour scale indicates the number of days since the start of the 2015
  outburst. Hardness ratio defined as $\mathrm{Flux_{(15-50keV)}/Flux_{(2-10keV)}}$ and
  intensity as $\mathrm{Flux_{(15-50keV)} + Flux_{(2-10keV)}}$.}\label{fig:4}

\end{figure*}

J1804 was detected in all 9 {\it Swift} observations which we used in
our study (Table~\ref{tab:obs}).  As can be seen from the full X-ray
light curve (see Fig. \ref{fig:3} and \citealt{Parikh2016} for more details),
the 2015 outburst of J1804 lasted $\sim 5$ months. The source was in the hard X-ray
state until 2015 March, with a typical ($0.5-10$\,keV) flux of $1.5 \times 10^{-9}$\,erg cm$^{-2}$s$^{-1}$,
before transitioning to the soft state around 3 April \citep{ATel_DEG2015} with a typical
flux of $\sim 4\times 10^{-9}$\,erg cm$^{-2}$s$^{-1}$ (see photon indices in Table~\ref{tab:obs}
and HID in Figure~\ref{fig:4}). Several weeks later (around June 6) the flux rapidly decreased as the source
returned to its quiescent level ($\sim 10^{-12}$--$10^{-13}$\,erg cm$^{-2}$s$^{-1}$; \citealt{Parikh2016}).

In order to investigate the X-ray/radio correlation, we estimated the
X-ray flux values corresponding to the exact times of our radio observations
by logarithmically interpolating between six pairs of closely spaced consecutive
X-ray observations (see blue circles in Figure \ref{fig:3}). In observation 1 we
found J1804 right among other NSs in the $L_R$/$L_X$ diagram.  In observation 2 we found
the radio emission to be strongly suppressed, providing the lowest soft-state radio
detection to date.  In observations 3--6 the source was not detected, which rules
out that it is as radio bright as tMSPs or AMXPs at low X-ray luminosities.

\section{Discussion}
\label{sec:disc}


\subsection{Jet quenching}


Jet quenching is a well-established phenomenon in BH-LMXBs. For most BH-LMXBs,
quenching of the radio emission is observed immediately before the transition
to the soft X-ray state (e.g. \citealt{Fen2009}). Only a few detections of very faint,
optically thin radio emission have been made in the soft state, immediately following
hard-to-soft X-ray state transitions (e.g. \citealt{Coriat2011}). Further into the
soft-state (between 5 to 30 days after the transition) and at lower X-ray luminosity,
numerous instances of radio non-detections indicate that the soft X-ray state radio
jet is heavily suppressed (\citealt{Hjellming1999}, \citealt{McClintock2009}, \citealt{Gallo2004}).

There are only three BH systems (XTE~J1650$-$500; \citealt{Corbel2004}, GX 339$-$4;
\citealt{Fen2009} and XTE~J1748$-$288; \citealt{Brocksopp2007}) which have been
detected in the full soft state (up to a month after the state transition). However,
the observed radio emission in these cases was optically thin and (in some cases) the
observed behaviour of spatially resolved components indicates that this emission is
most likely not from a compact jet, but rather from its remnant: i.e. the interaction
of ejected radio blobs with the surrounding interstellar medium or pre-existing jet
material \citep{Fen2016}. For most BH-LMXBs, radio emission is strongly quenched in
the soft state and reappears only after the system has transitioned back to the hard
state -- the most striking example being H1743$-$322, where the level of quenching was
detected to be a factor of more than 700 \citep{Coriat2011}. Soft-state jet quenching
(by a factor of $\gtrsim$25) was also observed in Swift~J1753.5$-$0127 \citep{Rushton2016}.
Furthermore, LMC~X-1 (\citealt{Gierlinski2001}) and 4U1957$+$11 (\citealt{Russell2011}),
which remain in a persistent soft state, do not show detectable radio emission. This is
consistent with the idea that a jet can not be produced in the soft X-ray state and the
emission that has been observed in other BH systems in this state could be just a remnant
of the transient ejecta.

The physical origin of this quenching is unknown since no widely accepted model of jet production
has been developed to date. It is thought that jet production can be further understood by
considering how the jet behaves during the different modes of accretion (i.e. X-ray states). 
The most typical models (see e.g. \citealt{Meier2001}) suggest that large-scale-height poloidal
magnetic fields are responsible for jet production. Indeed, in the hard X-ray state these fields
can be provided by a vertically extended accretion flow (i.e. the corona; see Section \ref{sec:intro}),
while at higher accretion rates, in the soft state, a geometrically-thin accretion disk alone
cannot provide such a magnetic field and thus cannot sustain a jet \citep{Meier2001}.  


Whether jet quenching also occurs in NS-LMXBs is currently unclear.
The handful of observed systems show markedly different behaviours.

\textbf{4U1728$-$34}

4U1728$-$34 was the first Atoll-type NS-LMXB that was well sampled in the 
X-ray/radio plane \citep{MIG2003}. This is a persistent source that
was monitored with two separate quasi-simultaneous X-ray/radio campaigns in 2000 and 2001.
During the 2000 campaign, 4U1728$-$34 stayed mostly in the hard X-ray state with two short
excursions into the soft state. The radio luminosities in the hard and soft states were almost
identical, indicating no radio quenching.  One observation (point `g' in \citealt{MIG2003} Fig.1,
or point `Q2' in \citealt{TUD2009} Fig.9) did provide a lower radio luminosity than other epochs,
however, its X-ray state was unclear (see `?' symbol in Figure~\ref{fig:1}). Therefore, jet
quenching cannot be definitively connected to an X-ray state in this case. During the second observational
campaign (2001), 4U1728$-$34 remained in the hard X-ray state with lower X-ray and radio
luminosities than observed in the previous campaign (when the system was transitioning
between hard and soft X-ray states). Combined observations from both campaigns, mixing hard and
soft states, revealed a $\beta = 1.4$ power-law correlation between the observed radio and X-ray
luminosities.

\textbf{Aql~X-1}

Aql~X-1 is one of the best studied NS-LMXBs, going into outburst
almost every year. It was found to have a tentative $\beta = 0.7$
power-law correlation between $L_R$ and $L_X$ for hard state observations
(\citealt{TUD2009}; \citealt{Tetarenko2016}; see green squares and green dotted line in
Figure~\ref{fig:1}). There were also several radio observations
of this source in its soft X-ray state: an upper limit in 2002, a
detection in 2004, and an upper limit and detection in 2009 (\citealt{MJ2010};
note that the upper limits were significantly deeper than the detections).
All observations that have been taken during the soft state appear
to be at a lower (up to a factor of 13) radio luminosity for a given $L_X$
than detections obtained during the hard state. Thus, Aql~X-1 was the
only NS-LMXB that showed a certain degree of jet quenching in the soft
X-ray state and at high X-ray luminosity \citep{MJ2010}. It should be noted that comparing hard and
soft radio flux densities from different outbursts of the same source can
in some cases be dangerous because NS-LMXBs can have very different
hardness-intensity tracks in each outburst (see e.g. \citealt{MD2014}).
However, over the outbursts in which the radio/X-ray correlation of
Aql~X-1 was compared, the observed HID tracks appeared very similar
(see \citealt{TUD2009}, \citealt{MJ2010}).

\textbf{Soft state only NS-LMXBs}

NS-LMXBs that have only been observed in the radio during a persistent soft state
appear to exhibit two distinct behaviours: a given source is either deeply quenched or
shows sustained radio emission. A single observation of GX~9+9 \citep{MIG2011} provided
a deep radio upper limit ($L_R \lesssim 4\times10^{27}$erg/s), indicating strong radio
quenching ($\ge$2 orders of magnitude when compared to typical hard state systems at
similar X-ray luminosity), similar to soft state BH systems. Ser~X-1 and 4U1820$-$30,
which are persistently in outburst \citep{MIG2004}, appear to show low levels of jet
quenching (being detected in radio at a slightly lower level). The few contemporaneous
radio and X-ray observations of these systems revealed around an order of magnitude
lower radio emission than what is expected for hard-state NSs by extrapolating the
$L_R/L_X$ correlation to higher $L_X$ \citep{MIGFEN2006}. However, X-ray monitoring
has shown that these systems have transitioned between the hard and soft
states, therefore, it is possible that the observed radio emission could be residual
emission from a pre-existing hard-state jet. Alternatively, the rapid burster MXB1730$-$335
\citep{Moore2000} did not appear to show any jet quenching during its soft X-ray outburst\footnote{We analysed publicly available {\it RXTE} data from 6 and 11 November 1996 (when the radio obervations of \citet{Moore2000} were taken).
We found strong blackbody (with temperature $\sim 1-3$ keV) and very steep powerlaw ($\Gamma > 3$) components in both spectra.
This confirms that radio observations of MXB1730$-$335 were taken while it was indeed in the soft X-ray state.}.
However, hard-state radio emission has not been observed in any of these three systems and
their perceived level of jet quenching has been assumed by comparing their soft-state radio
emission to the assumed $\beta=1.4$ relation for typical hard-state systems. Therefore, we can not determine their true level of
jet quenching as reliably as observing hard-to-soft state transitions during the same outburst.

\textbf{1RXS~J180408.9$-$342058}


Observations during both hard and soft X-ray states for a single outburst of the same
source are critical to accurately track the evolution of the radio jet and to measure its quenching. So far this has only been achieved for Aql~X-1, which showed
good evidence for jet quenching \citep{MJ2010}. Thus, J1804 is only the second source
in which we can examine NS-LMXB jet quenching in reasonable detail.


We performed radio observations of J1804 during outburst and its evolution back to
quiescence (see HID in Figure~{\ref{fig:4}}).  During the outburst, the source was
detected at radio wavelengths in both hard and soft X-ray states (see Figure~\ref{fig:3}).
We detected J1804 in the rising hard state, finding that its radio and X-ray luminosities
were consistent with typical Atoll-type NS-LMXB hard state luminosities. We also significantly
detected J1804 in the soft X-ray state. This detection is a factor of $\sim\, 12$ lower in
flux density than the hard state detection ($\sim$1 month earlier), hence we can firmly
claim the occurrence of radio jet quenching by more than an order of magnitude.

 
The radio spectrum in our first observational epoch was consistent with being flat
($\alpha = 0.12 \pm 0.18$), suggesting optically-thick synchrotron emission from
a compact jet. During the soft X-ray state, the radio source was faint ($\sim 19 \pm 4\, \mu$Jy;
$\sim 5\sigma$ detection), which precludes any detailed spectral analysis for this observation.
The obtained spectral index for this epoch ($\alpha = -0.31 \pm 0.68$) has a large uncertainty 
and thus does not discriminate between optically thick and optically thin synchrotron emission
from a radio jet ($\alpha \approx$0).
This detection is at a lower radio luminosity than any other detected NS-LMXB in outburst\footnote{Note that
one observation of the tMSP PSR~J1023+0038 (the lowest triangle point) has a lower radio luminosity
\citep{DEL2015}, owing to its close proximity ($\sim$1 kpc).  Likewise, the next tMSP triangle point to the
right in Figure~\ref{fig:1}, which is XSS~J12270$-$4859 \citep{Hill2011}}. Although GX~9+9 has a
similar radio luminosity, this is only an upper limit (see \citealt{MIG2011}). 


Since we only have one radio observation during the hard state, and one during the soft
state (with $\sim$ 1 month separation), it could be argued that the soft-state detection
represents residual emission of the hard-state jet (although it is 10 days after the hard-to-soft state transition). Therefore, the true level of quenching could, in principle, be deeper. Remnant radio
emission has been seen in BH systems during the soft state (see e.g. \citealt{Fen2009}
and references therein). In this case, the collimated jet is usually quenched prior to the X-ray
state transition and a transient jet is launched during the transition. Therefore, we can not
rule out the possibility that both of these events occurred between our first and second radio
observations of J1804 and that the observed radio emission in the soft state is residual emission
from a transient jet. If this was the case, our soft state detection would be at least a lower limit
on the true level of the jet quenching.

NS-LMXBs have shown a broad diversity of behaviour, and only with
observations of more such systems, over the widest possible range of
radio and X-ray luminosities, can we hope to ascertain whether they
show a consistent set of behaviours and make detailed comparisons to
BH-LMXBs. Some NS-LMXBs appear to quench in the soft state (e.g. Aql~X-1,
GX~9+9 and J1804) and others apparently do not (4U1748$-$34, Ser~X-1, 4U1820$-$30
and MXB1730$-$335). Therefore, there may exist some tunable parameter that
affects jet production and quenching, such as magnetic field strength,
spin of the NS and/or mass accretion rate. \citealt{MIG2011} summarised
observations of all NS-LMXBs observed in the radio band to determine a unified
scheme of jet production based on such parameters, however, they could
not identify any evidence for a single parameter being responsible.

Our soft-state radio detection was only possible due to the recent upgrade
to the VLA \citep{VLA2011}, which increased its sensitivity by a factor of
$\sim\, 10$ for similar observations to those here. This increased sensitivity
opens up the possibility to study very faint radio sources such as those associated
with jet emission in NS-LMXBs. Further increased sensitivity would allow us to study
the spectral properties of the radio emission from NS-LMXBs in detail (e.g. distinguishing between optically thin and thick emission) and possibly detect even
deeper radio quenching.
 
\subsection{Radio/X-ray correlation}

We observed J1804 in the radio during its 2015 outburst, at high X-ray luminosity $L_X > 10^{36}$ erg/s in both
hard and soft X-ray states, and at lower X-ray luminosity as it returned towards quiescence (see Figure~\ref{fig:1}).
The source was not detected below $L_X < 10^{35}$\, erg/s. Combining
a hard-state detection with our quiescent limits, we conclude that J1804
follows a $L_R/L_X$ track with an index of $\beta \gtrsim 0.7$,
in agreement with typical non-pulsating NS-LMXBs\footnote{Here we make a distinction between pulsating
NS-LMXBs, which show coherent X-ray pulsations tied to the NS rotation rate, and non-pulsating NS-LMXBs,
which do not}, for which $\beta \approx 1.4$ \citep{Tetarenko2016}.

Pulsating NS-LMXBs, such as AMXPs and tMSPs, appear to be more radio bright
at low X-ray luminosity ($L_X \sim 10^{32-35}$ erg/s) than non-pulsating
systems (see magenta stars and dark blue triangles in Figure~\ref{fig:1},
$L_X \sim 10^{32-35}$ erg/s range). However, at higher X-ray luminosities
($L_X > 10^{36}$ erg/s), some AMXPs show similar (e.g. radio observations of
SAX~J1808.4$-$3658; \citealt{Rupen2002}, \citealt{MIGFEN2006}) or even lower radio
luminosity (see \citealt{Tudor2017}) compared with non-pulsating systems (assuming that the relative distances are correct).
Furthermore, there are three confirmed tMSP systems observed so far: PSR~J1023+0038 \citep{Stappers2014} and
XSS~J12270$-$4859 \citep{Bassa2014} have so far remained in a very low and steady accretion state,
and IGR~J18245$-$2452 (M28I; \citealt{Papitto2013}) also went into full outburst.
As suggested by \citealt{DEL2015}, these systems possibly obey their own radio/X-ray
correlation with $\beta \approx 0.6$ (see the dark blue dashed line in Figure~\ref{fig:1}).
This was suggested by connecting the highest point (M28I) with the two low-accretion-rate systems
whose radio brightness is markedly higher than what is expected for non-pulsating NS-LMXBs
based on a $\beta = 1.4$ power law correlation.

This difference opens up the possibility to distinguish between different classes of accreting NSs using
their radio/X-ray behaviour. However, due to the expectation that they would be too faint to detect,
very few NS-LMXBs have been targeted in the radio at low X-ray luminosities ($L_X \sim\, 10^{32}-10^{35}$\,erg/s).
One of the few exceptions is the recent non-detection of EXO~1745$-$248 \citep{Tetarenko2016} at an X-ray
luminosity of $\sim 10^{35}$ erg/s, which indicates that this system is significantly radio fainter compared with
the tMSPs. Lastly, the upper limits provided by J1804 also indicate that the radio emission of some
NS-LMXBs at low X-ray luminosity is much less (by at least a factor of a few and possibly by an order of magnitude or more)
than that of tMSPs and AMXPs (\citealt{DEL2015}, \citealt{Tudor2017}).

\section{Conclusion}

We find significant radio jet quenching in the non-pulsating NS-LMXB J1804 (by a factor of $\sim$12) during its transition from hard to soft X-ray states in its 2015 outburst.
Such jet quenching has been previously observed, arguably, only in two other NS-LMXBs: Aql~X-1, where quasi-simultaneous radio/X-ray data tracked both the hard and soft X-ray states during two outbursts and GX~9+9, where a single upper limit in the soft state is the only available evidence.
Other NS-LMXBs studied quasi-simultaneously in radio/X-ray have not clearly shown jet quenching.
Furthermore, our radio observations of J1804 at $L_X \sim\, 10^{35}$ erg/s seem to
indicate that it has a lower radio luminosity than previously observed pulsating NS-LMXBs, i.e. the tMSPs and AMXPs. Such behaviour has also been
observed in the non-pulsating NS-LMXB EXO~1745$-$245, which likewise shows only a radio upper-limit at $L_X \sim\, 10^{35}$ erg/s. 

Our results indicate that NS-LMXBs have fundamentally different accretion characteristics within the class.
However, only a few systems have been observed in a broad X-ray/radio luminosity range and during various X-ray states.
Observations of more systems at $L_X<10^{36}$ erg/s are needed to characterize this phenomenon.


\section*{Acknowledgements}

NVG acknowledges funding from NOVA. ATD is the recipient of an Australian
Research Council Future Fellowship (FT150100415).  JWTH acknowledges funding
from an NWO Vidi fellowship and from the European Research Council
under the European Union's Seventh Framework Programme (FP/2007-2013)
/ ERC Starting Grant agreement nr. 337062 (``DRAGNET''). 
JCAMJ is the recipient of an Australian Research Council Future Fellowship
(FT140101082). ND is supported by a Vidi grant from NWO and a Marie Curie
fellowship (FP-PEOPLE-2013-IEF-627148) from the European Commission. 
RW and ASP are supported by a NWO Top Grant, Module 1, awarded to RW.
TDR acknowledges support from the Netherlands Organisation for Scientific
Research (NWO) Veni Fellowship, grant number 639.041.646. DA acknowledges
support from the Royal Society. We are grateful to Neil Gehrels and the duty
scientists for rapid scheduling of the {\it Swift} observations. We acknowledge
the use of the {\it Swift} public data archive.

\bibliographystyle{mnras}
\bibliography{allbib}

\begin{thebibliography}{}
\makeatletter
\relax
\def\mn@urlcharsother{\let\do\@makeother \do\$\do\&\do\#\do\^\do\_\do\%\do\~}
\def\mn@doi{\begingroup\mn@urlcharsother \@ifnextchar [ {\mn@doi@}
  {\mn@doi@[]}}
\def\mn@doi@[#1]#2{\def\@tempa{#1}\ifx\@tempa\@empty \href
  {http://dx.doi.org/#2} {doi:#2}\else \href {http://dx.doi.org/#2} {#1}\fi
  \endgroup}
\def\mn@eprint#1#2{\mn@eprint@#1:#2::\@nil}
\def\mn@eprint@arXiv#1{\href {http://arxiv.org/abs/#1} {{\tt arXiv:#1}}}
\def\mn@eprint@dblp#1{\href {http://dblp.uni-trier.de/rec/bibtex/#1.xml}
  {dblp:#1}}
\def\mn@eprint@#1:#2:#3:#4\@nil{\def\@tempa {#1}\def\@tempb {#2}\def\@tempc
  {#3}\ifx \@tempc \@empty \let \@tempc \@tempb \let \@tempb \@tempa \fi \ifx
  \@tempb \@empty \def\@tempb {arXiv}\fi \@ifundefined
  {mn@eprint@\@tempb}{\@tempb:\@tempc}{\expandafter \expandafter \csname
  mn@eprint@\@tempb\endcsname \expandafter{\@tempc}}}

\bibitem[\protect\citeauthoryear{{Baglio}, {Campana}  \& {D'Avanzo}}{{Baglio}
  et~al.}{2015}]{ATel_BAG2015}
{Baglio} M.~C.,  {Campana} S.,   {D'Avanzo} P.,  2015, The Astronomer's
  Telegram, \href {http://adsabs.harvard.edu/abs/2015ATel.7100....1B} {7100}

\bibitem[\protect\citeauthoryear{{Baglio}, {D'Avanzo}, {Campana}, {Goldoni},
  {Masetti}, {Mu{\~n}oz-Darias}, {Pati{\~n}o-{\'A}lvarez}  \&
  {Chavushyan}}{{Baglio} et~al.}{2016}]{BAG2016}
{Baglio} M.~C.,  {D'Avanzo} P.,  {Campana} S.,  {Goldoni} P.,  {Masetti} N.,
  {Mu{\~n}oz-Darias} T.,  {Pati{\~n}o-{\'A}lvarez} V.,   {Chavushyan} V.,
  2016, \mn@doi [\aap] {10.1051/0004-6361/201527147}, \href
  {http://adsabs.harvard.edu/abs/2016A%26A...587A.102B} {587, A102}

\bibitem[\protect\citeauthoryear{{Bassa} et~al.,}{{Bassa}
  et~al.}{2014}]{Bassa2014}
{Bassa} C.~G.,  et~al., 2014, \mn@doi [\mnras] {10.1093/mnras/stu708}, \href
  {http://adsabs.harvard.edu/abs/2014MNRAS.441.1825B} {441, 1825}

\bibitem[\protect\citeauthoryear{{Begelman}, {McKee}  \& {Shields}}{{Begelman}
  et~al.}{1983}]{Begelman1983}
{Begelman} M.~C.,  {McKee} C.~F.,   {Shields} G.~A.,  1983, \mn@doi [\apj]
  {10.1086/161178}, \href {http://adsabs.harvard.edu/abs/1983ApJ...271...70B}
  {271, 70}

\bibitem[\protect\citeauthoryear{{Bogdanov} et~al.,}{{Bogdanov}
  et~al.}{2015}]{Bogdanov2015}
{Bogdanov} S.,  et~al., 2015, \mn@doi [\apj] {10.1088/0004-637X/806/2/148},
  \href {http://adsabs.harvard.edu/abs/2015ApJ...806..148B} {806, 148}

\bibitem[\protect\citeauthoryear{{Boissay} et~al.,}{{Boissay}
  et~al.}{2015}]{INTEGRAL2015_ATel}
{Boissay} R.,  et~al., 2015, The Astronomer's Telegram, \href
  {http://adsabs.harvard.edu/abs/2015ATel.7096....1B} {7096}

\bibitem[\protect\citeauthoryear{{Brocksopp}, {Miller-Jones}, {Fender}  \&
  {Stappers}}{{Brocksopp} et~al.}{2007}]{Brocksopp2007}
{Brocksopp} C.,  {Miller-Jones} J.~C.~A.,  {Fender} R.~P.,   {Stappers} B.~W.,
  2007, \mn@doi [\mnras] {10.1111/j.1365-2966.2007.11846.x}, \href
  {http://adsabs.harvard.edu/abs/2007MNRAS.378.1111B} {378, 1111}

\bibitem[\protect\citeauthoryear{{Chenevez} et~al.,}{{Chenevez}
  et~al.}{2012}]{ATel_INTEGRAL2012}
{Chenevez} J.,  et~al., 2012, The Astronomer's Telegram, \href
  {http://adsabs.harvard.edu/abs/2012ATel.4050....1C} {4050}

\bibitem[\protect\citeauthoryear{{Condon}, {Cotton}, {Greisen}, {Yin},
  {Perley}, {Taylor}  \& {Broderick}}{{Condon} et~al.}{1998}]{CON1998}
{Condon} J.~J.,  {Cotton} W.~D.,  {Greisen} E.~W.,  {Yin} Q.~F.,  {Perley}
  R.~A.,  {Taylor} G.~B.,   {Broderick} J.~J.,  1998, \mn@doi [\aj]
  {10.1086/300337}, \href {http://adsabs.harvard.edu/abs/1998AJ....115.1693C}
  {115, 1693}

\bibitem[\protect\citeauthoryear{{Corbel}, {Fender}, {Tzioumis}, {Nowak},
  {McIntyre}, {Durouchoux}  \& {Sood}}{{Corbel} et~al.}{2000}]{Corbel2000}
{Corbel} S.,  {Fender} R.~P.,  {Tzioumis} A.~K.,  {Nowak} M.,  {McIntyre} V.,
  {Durouchoux} P.,   {Sood} R.,  2000, \aap, \href
  {http://adsabs.harvard.edu/abs/2000A%26A...359..251C} {359, 251}

\bibitem[\protect\citeauthoryear{{Corbel}, {Fender}, {Tomsick}, {Tzioumis}  \&
  {Tingay}}{{Corbel} et~al.}{2004}]{Corbel2004}
{Corbel} S.,  {Fender} R.~P.,  {Tomsick} J.~A.,  {Tzioumis} A.~K.,   {Tingay}
  S.,  2004, \mn@doi [\apj] {10.1086/425650}, \href
  {http://adsabs.harvard.edu/abs/2004ApJ...617.1272C} {617, 1272}

\bibitem[\protect\citeauthoryear{{Coriat} et~al.,}{{Coriat}
  et~al.}{2011}]{Coriat2011}
{Coriat} M.,  et~al., 2011, \mn@doi [\mnras]
  {10.1111/j.1365-2966.2011.18433.x}, \href
  {http://adsabs.harvard.edu/abs/2011MNRAS.414..677C} {414, 677}

\bibitem[\protect\citeauthoryear{{Coriat}, {Fender}  \& {Dubus}}{{Coriat}
  et~al.}{2012}]{Coriat2012}
{Coriat} M.,  {Fender} R.~P.,   {Dubus} G.,  2012, \mn@doi [\mnras]
  {10.1111/j.1365-2966.2012.21339.x}, \href
  {http://adsabs.harvard.edu/abs/2012MNRAS.424.1991C} {424, 1991}

\bibitem[\protect\citeauthoryear{{Degenaar} et~al.,}{{Degenaar}
  et~al.}{2014}]{Deg2014}
{Degenaar} N.,  et~al., 2014, \mn@doi [\apj] {10.1088/0004-637X/792/2/109},
  \href {http://adsabs.harvard.edu/abs/2014ApJ...792..109D} {792, 109}

\bibitem[\protect\citeauthoryear{{Degenaar} et~al.,}{{Degenaar}
  et~al.}{2015}]{ATel_DEG2015}
{Degenaar} N.,  et~al., 2015, The Astronomer's Telegram, \href
  {http://adsabs.harvard.edu/abs/2015ATel.7352....1D} {7352}

\bibitem[\protect\citeauthoryear{{Degenaar} et~al.,}{{Degenaar}
  et~al.}{2016}]{Deg2016}
{Degenaar} N.,  et~al., 2016, \mn@doi [\mnras] {10.1093/mnras/stw1593}, \href
  {http://adsabs.harvard.edu/abs/2016MNRAS.461.4049D} {461, 4049}

\bibitem[\protect\citeauthoryear{{Deller} et~al.,}{{Deller}
  et~al.}{2015a}]{DEL2015}
{Deller} A.~T.,  et~al., 2015a, \mn@doi [\apj] {10.1088/0004-637X/809/1/13},
  \href {http://adsabs.harvard.edu/abs/2015ApJ...809...13D} {809, 13}

\bibitem[\protect\citeauthoryear{{Deller} et~al.,}{{Deller}
  et~al.}{2015b}]{ATel_DEL2015}
{Deller} A.,  et~al., 2015b, The Astronomer's Telegram, \href
  {http://adsabs.harvard.edu/abs/2015ATel.7255....1D} {7255}

\bibitem[\protect\citeauthoryear{{Esin}, {McClintock}  \& {Narayan}}{{Esin}
  et~al.}{1997}]{Esin1997}
{Esin} A.~A.,  {McClintock} J.~E.,   {Narayan} R.,  1997, \apj, \href
  {http://adsabs.harvard.edu/abs/1997ApJ...489..865E} {489, 865}

\bibitem[\protect\citeauthoryear{{Falcke}, {K{\"o}rding}  \&
  {Markoff}}{{Falcke} et~al.}{2004}]{Falcke2004}
{Falcke} H.,  {K{\"o}rding} E.,   {Markoff} S.,  2004, \mn@doi [\aap]
  {10.1051/0004-6361:20031683}, \href
  {http://adsabs.harvard.edu/abs/2004A%26A...414..895F} {414, 895}

\bibitem[\protect\citeauthoryear{{Fender} \& {Kuulkers}}{{Fender} \&
  {Kuulkers}}{2001}]{FENKUUL2001}
{Fender} R.~P.,  {Kuulkers} E.,  2001, \mn@doi [\mnras]
  {10.1046/j.1365-8711.2001.04345.x}, \href
  {http://adsabs.harvard.edu/abs/2001MNRAS.324..923F} {324, 923}

\bibitem[\protect\citeauthoryear{{Fender} \& {Mu{\~n}oz-Darias}}{{Fender} \&
  {Mu{\~n}oz-Darias}}{2016}]{Fen2016}
{Fender} R.,  {Mu{\~n}oz-Darias} T.,  2016, in {Haardt} F.,  {Gorini} V.,
  {Moschella} U.,  {Treves} A.,   {Colpi} M.,  eds,  Lecture Notes in Physics,
  Berlin Springer Verlag Vol. 905, Lecture Notes in Physics, Berlin Springer
  Verlag. p.~65 (\mn@eprint {arXiv} {1505.03526}),
  \mn@doi{10.1007/978-3-319-19416-5_3}

\bibitem[\protect\citeauthoryear{{Fender} et~al.,}{{Fender}
  et~al.}{1999}]{Fen1999}
{Fender} R.,  et~al., 1999, \mn@doi [\apjl] {10.1086/312128}, \href
  {http://adsabs.harvard.edu/abs/1999ApJ...519L.165F} {519, L165}

\bibitem[\protect\citeauthoryear{{Fender}, {Gallo}  \& {Jonker}}{{Fender}
  et~al.}{2003}]{FenGalJon2003}
{Fender} R.~P.,  {Gallo} E.,   {Jonker} P.~G.,  2003, \mn@doi [\mnras]
  {10.1046/j.1365-8711.2003.06950.x}, \href
  {http://adsabs.harvard.edu/abs/2003MNRAS.343L..99F} {343, L99}

\bibitem[\protect\citeauthoryear{{Fender}, {Belloni}  \& {Gallo}}{{Fender}
  et~al.}{2004a}]{FENBELGAL2004}
{Fender} R.~P.,  {Belloni} T.~M.,   {Gallo} E.,  2004a, \mn@doi [\mnras]
  {10.1111/j.1365-2966.2004.08384.x}, \href
  {http://adsabs.harvard.edu/abs/2004MNRAS.355.1105F} {355, 1105}

\bibitem[\protect\citeauthoryear{{Fender}, {Belloni}  \& {Gallo}}{{Fender}
  et~al.}{2004b}]{Fen2004}
{Fender} R.~P.,  {Belloni} T.~M.,   {Gallo} E.,  2004b, \mn@doi [\mnras]
  {10.1111/j.1365-2966.2004.08384.x}, \href
  {http://adsabs.harvard.edu/abs/2004MNRAS.355.1105F} {355, 1105}

\bibitem[\protect\citeauthoryear{{Fender}, {Homan}  \& {Belloni}}{{Fender}
  et~al.}{2009}]{Fen2009}
{Fender} R.~P.,  {Homan} J.,   {Belloni} T.~M.,  2009, \mn@doi [\mnras]
  {10.1111/j.1365-2966.2009.14841.x}, \href
  {http://adsabs.harvard.edu/abs/2009MNRAS.396.1370F} {396, 1370}

\bibitem[\protect\citeauthoryear{{Gallo}, {Fender}  \& {Pooley}}{{Gallo}
  et~al.}{2003}]{GALLO2003}
{Gallo} E.,  {Fender} R.~P.,   {Pooley} G.~G.,  2003, \mn@doi [\mnras]
  {10.1046/j.1365-8711.2003.06791.x}, \href
  {http://adsabs.harvard.edu/abs/2003MNRAS.344...60G} {344, 60}

\bibitem[\protect\citeauthoryear{{Gallo}, {Corbel}, {Fender}, {Maccarone}  \&
  {Tzioumis}}{{Gallo} et~al.}{2004}]{Gallo2004}
{Gallo} E.,  {Corbel} S.,  {Fender} R.~P.,  {Maccarone} T.~J.,   {Tzioumis}
  A.~K.,  2004, \mn@doi [\mnras] {10.1111/j.1365-2966.2004.07435.x}, \href
  {http://adsabs.harvard.edu/abs/2004MNRAS.347L..52G} {347, L52}

\bibitem[\protect\citeauthoryear{{Gallo}, {Fender}, {Miller-Jones}, {Merloni},
  {Jonker}, {Heinz}, {Maccarone}  \& {van der Klis}}{{Gallo}
  et~al.}{2006}]{Gallo2006}
{Gallo} E.,  {Fender} R.~P.,  {Miller-Jones} J.~C.~A.,  {Merloni} A.,  {Jonker}
  P.~G.,  {Heinz} S.,  {Maccarone} T.~J.,   {van der Klis} M.,  2006, \mn@doi
  [\mnras] {10.1111/j.1365-2966.2006.10560.x}, \href
  {http://adsabs.harvard.edu/abs/2006MNRAS.370.1351G} {370, 1351}

\bibitem[\protect\citeauthoryear{{Gallo}, {Miller}  \& {Fender}}{{Gallo}
  et~al.}{2012}]{Gallo2012}
{Gallo} E.,  {Miller} B.~P.,   {Fender} R.,  2012, \mn@doi [\mnras]
  {10.1111/j.1365-2966.2012.20899.x}, \href
  {http://adsabs.harvard.edu/abs/2012MNRAS.423..590G} {423, 590}

\bibitem[\protect\citeauthoryear{{Gallo} et~al.,}{{Gallo}
  et~al.}{2014}]{Gallo2014}
{Gallo} E.,  et~al., 2014, \mn@doi [\mnras] {10.1093/mnras/stu1599}, \href
  {http://adsabs.harvard.edu/abs/2014MNRAS.445..290G} {445, 290}

\bibitem[\protect\citeauthoryear{{Gierli{\'n}ski},
  {Macio{\l}ek-Nied{\'z}wiecki}  \& {Ebisawa}}{{Gierli{\'n}ski}
  et~al.}{2001}]{Gierlinski2001}
{Gierli{\'n}ski} M.,  {Macio{\l}ek-Nied{\'z}wiecki} A.,   {Ebisawa} K.,  2001,
  \mn@doi [\mnras] {10.1046/j.1365-8711.2001.04540.x}, \href
  {http://adsabs.harvard.edu/abs/2001MNRAS.325.1253G} {325, 1253}

\bibitem[\protect\citeauthoryear{{Hasinger} \& {van der Klis}}{{Hasinger} \&
  {van der Klis}}{1989}]{HV1989}
{Hasinger} G.,  {van der Klis} M.,  1989, \aap, \href
  {http://adsabs.harvard.edu/abs/1989A%26A...225...79H} {225, 79}

\bibitem[\protect\citeauthoryear{{Heinke}, {Bahramian}, {Degenaar}  \&
  {Wijnands}}{{Heinke} et~al.}{2015}]{Heinke2015}
{Heinke} C.~O.,  {Bahramian} A.,  {Degenaar} N.,   {Wijnands} R.,  2015,
  \mn@doi [\mnras] {10.1093/mnras/stu2652}, \href
  {http://adsabs.harvard.edu/abs/2015MNRAS.447.3034H} {447, 3034}

\bibitem[\protect\citeauthoryear{{Hill} et~al.,}{{Hill}
  et~al.}{2011}]{Hill2011}
{Hill} A.~B.,  et~al., 2011, \mn@doi [\mnras]
  {10.1111/j.1365-2966.2011.18692.x}, \href
  {http://adsabs.harvard.edu/abs/2011MNRAS.415..235H} {415, 235}

\bibitem[\protect\citeauthoryear{{Hjellming} et~al.,}{{Hjellming}
  et~al.}{1999}]{Hjellming1999}
{Hjellming} R.~M.,  et~al., 1999, \mn@doi [\apj] {10.1086/306948}, \href
  {http://adsabs.harvard.edu/abs/1999ApJ...514..383H} {514, 383}

\bibitem[\protect\citeauthoryear{{Homan} et~al.,}{{Homan}
  et~al.}{2010}]{Homan2010}
{Homan} J.,  et~al., 2010, \mn@doi [\apj] {10.1088/0004-637X/719/1/201}, \href
  {http://adsabs.harvard.edu/abs/2010ApJ...719..201H} {719, 201}

\bibitem[\protect\citeauthoryear{{Homan}, {Fridriksson}, {Wijnands}, {Cackett},
  {Degenaar}, {Linares}, {Lin}  \& {Remillard}}{{Homan}
  et~al.}{2014}]{Homan2014}
{Homan} J.,  {Fridriksson} J.~K.,  {Wijnands} R.,  {Cackett} E.~M.,  {Degenaar}
  N.,  {Linares} M.,  {Lin} D.,   {Remillard} R.~A.,  2014, \mn@doi [\apj]
  {10.1088/0004-637X/795/2/131}, \href
  {http://adsabs.harvard.edu/abs/2014ApJ...795..131H} {795, 131}

\bibitem[\protect\citeauthoryear{{Kaur} \& {Heinke}}{{Kaur} \&
  {Heinke}}{2012}]{ATel_Quiescence2012}
{Kaur} R.,  {Heinke} C.,  2012, The Astronomer's Telegram, \href
  {http://adsabs.harvard.edu/abs/2012ATel.4085....1K} {4085}

\bibitem[\protect\citeauthoryear{{Krimm} et~al.,}{{Krimm}
  et~al.}{2015a}]{ATel_Swift2015_detecion}
{Krimm} H.~A.,  et~al., 2015a, The Astronomer's Telegram, \href
  {http://adsabs.harvard.edu/abs/2015ATel.6997....1K} {6997}

\bibitem[\protect\citeauthoryear{{Krimm}, {Kennea}, {Siegel}  \&
  {Sbarufatti}}{{Krimm} et~al.}{2015b}]{ATel_Swift2015}
{Krimm} H.~A.,  {Kennea} J.~A.,  {Siegel} M.~H.,   {Sbarufatti} B.,  2015b, The
  Astronomer's Telegram, \href
  {http://adsabs.harvard.edu/abs/2015ATel.7039....1K} {7039}

\bibitem[\protect\citeauthoryear{{Kuulkers}, {den Hartog}, {in't Zand},
  {Verbunt}, {Harris}  \& {Cocchi}}{{Kuulkers} et~al.}{2003}]{Kuulkers2003}
{Kuulkers} E.,  {den Hartog} P.~R.,  {in't Zand} J.~J.~M.,  {Verbunt} F.~W.~M.,
   {Harris} W.~E.,   {Cocchi} M.,  2003, \mn@doi [\aap]
  {10.1051/0004-6361:20021781}, \href
  {http://adsabs.harvard.edu/abs/2003A%26A...399..663K} {399, 663}

\bibitem[\protect\citeauthoryear{{Ludlam} et~al.,}{{Ludlam}
  et~al.}{2016}]{Ludlam2016}
{Ludlam} R.~M.,  et~al., 2016, \mn@doi [\apj] {10.3847/0004-637X/824/1/37},
  \href {http://adsabs.harvard.edu/abs/2016ApJ...824...37L} {824, 37}

\bibitem[\protect\citeauthoryear{{Maccarone}}{{Maccarone}}{2012}]{Maccarone2012}
{Maccarone} T.~J.,  2012, preprint, \href
  {http://adsabs.harvard.edu/abs/2012arXiv1204.3154M} {} (\mn@eprint {arXiv}
  {1204.3154})

\bibitem[\protect\citeauthoryear{{Malzac}}{{Malzac}}{2007}]{Malzac2007}
{Malzac} J.,  2007, \mn@doi [\apss] {10.1007/s10509-007-9558-9}, \href
  {http://adsabs.harvard.edu/abs/2007Ap%26SS.311..149M} {311, 149}

\bibitem[\protect\citeauthoryear{{Matsuoka} et~al.,}{{Matsuoka}
  et~al.}{2009}]{Matsuoka2009}
{Matsuoka} M.,  et~al., 2009, \mn@doi [\pasj] {10.1093/pasj/61.5.999}, \href
  {http://adsabs.harvard.edu/abs/2009PASJ...61..999M} {61, 999}

\bibitem[\protect\citeauthoryear{{McClintock} \& {Remillard}}{{McClintock} \&
  {Remillard}}{2006}]{McC2004}
{McClintock} J.~E.,  {Remillard} R.~A.,  2006, {Black hole binaries}.
pp 157--213

\bibitem[\protect\citeauthoryear{{McClintock}, {Remillard}, {Rupen}, {Torres},
  {Steeghs}, {Levine}  \& {Orosz}}{{McClintock} et~al.}{2009}]{McClintock2009}
{McClintock} J.~E.,  {Remillard} R.~A.,  {Rupen} M.~P.,  {Torres} M.~A.~P.,
  {Steeghs} D.,  {Levine} A.~M.,   {Orosz} J.~A.,  2009, \mn@doi [\apj]
  {10.1088/0004-637X/698/2/1398}, \href
  {http://adsabs.harvard.edu/abs/2009ApJ...698.1398M} {698, 1398}

\bibitem[\protect\citeauthoryear{{McMullin}, {Waters}, {Schiebel}, {Young}  \&
  {Golap}}{{McMullin} et~al.}{2007}]{CASA2007}
{McMullin} J.~P.,  {Waters} B.,  {Schiebel} D.,  {Young} W.,   {Golap} K.,
  2007, in {Shaw} R.~A.,  {Hill} F.,   {Bell} D.~J.,  eds,  Astronomical
  Society of the Pacific Conference Series Vol. 376, Astronomical Data Analysis
  Software and Systems XVI. p.~127

\bibitem[\protect\citeauthoryear{{Meier}}{{Meier}}{2001}]{Meier2001}
{Meier} D.~L.,  2001, \mn@doi [\apjl] {10.1086/318921}, \href
  {http://adsabs.harvard.edu/abs/2001ApJ...548L...9M} {548, L9}

\bibitem[\protect\citeauthoryear{{Merloni}, {Heinz}  \& {di Matteo}}{{Merloni}
  et~al.}{2003}]{Merloni2003}
{Merloni} A.,  {Heinz} S.,   {di Matteo} T.,  2003, \mn@doi [\mnras]
  {10.1046/j.1365-2966.2003.07017.x}, \href
  {http://adsabs.harvard.edu/abs/2003MNRAS.345.1057M} {345, 1057}

\bibitem[\protect\citeauthoryear{{Meyer-Hofmeister} \&
  {Meyer}}{{Meyer-Hofmeister} \& {Meyer}}{2014}]{MeyerHof2014}
{Meyer-Hofmeister} E.,  {Meyer} F.,  2014, \mn@doi [\aap]
  {10.1051/0004-6361/201322423}, \href
  {http://adsabs.harvard.edu/abs/2014A%26A...562A.142M} {562, A142}

\bibitem[\protect\citeauthoryear{{Migliari}}{{Migliari}}{2011}]{MIG2011}
{Migliari} S.,  2011, in {Romero} G.~E.,  {Sunyaev} R.~A.,   {Belloni} T.,
  eds,  IAU Symposium Vol. 275, Jets at All Scales. pp 233--241,
  \mn@doi{10.1017/S174392131001608X}

\bibitem[\protect\citeauthoryear{{Migliari} \& {Fender}}{{Migliari} \&
  {Fender}}{2006}]{MIGFEN2006}
{Migliari} S.,  {Fender} R.~P.,  2006, \mn@doi [\mnras]
  {10.1111/j.1365-2966.2005.09777.x}, \href
  {http://adsabs.harvard.edu/abs/2006MNRAS.366...79M} {366, 79}

\bibitem[\protect\citeauthoryear{{Migliari}, {Fender}, {Rupen}, {Jonker},
  {Klein-Wolt}, {Hjellming}  \& {van der Klis}}{{Migliari}
  et~al.}{2003}]{MIG2003}
{Migliari} S.,  {Fender} R.~P.,  {Rupen} M.,  {Jonker} P.~G.,  {Klein-Wolt} M.,
   {Hjellming} R.~M.,   {van der Klis} M.,  2003, \mn@doi [\mnras]
  {10.1046/j.1365-8711.2003.06795.x}, \href
  {http://adsabs.harvard.edu/abs/2003MNRAS.342L..67M} {342, L67}

\bibitem[\protect\citeauthoryear{{Migliari}, {Fender}, {Rupen}, {Wachter},
  {Jonker}, {Homan}  \& {van der Klis}}{{Migliari} et~al.}{2004}]{MIG2004}
{Migliari} S.,  {Fender} R.~P.,  {Rupen} M.,  {Wachter} S.,  {Jonker} P.~G.,
  {Homan} J.,   {van der Klis} M.,  2004, \mn@doi [\mnras]
  {10.1111/j.1365-2966.2004.07768.x}, \href
  {http://adsabs.harvard.edu/abs/2004MNRAS.351..186M} {351, 186}

\bibitem[\protect\citeauthoryear{{Migliari}, {Fender}  \& {van der
  Klis}}{{Migliari} et~al.}{2005}]{MIG2005}
{Migliari} S.,  {Fender} R.~P.,   {van der Klis} M.,  2005, \mn@doi [\mnras]
  {10.1111/j.1365-2966.2005.09412.x}, \href
  {http://adsabs.harvard.edu/abs/2005MNRAS.363..112M} {363, 112}

\bibitem[\protect\citeauthoryear{{Miller-Jones} et~al.,}{{Miller-Jones}
  et~al.}{2010}]{MJ2010}
{Miller-Jones} J.~C.~A.,  et~al., 2010, \mn@doi [\apjl]
  {10.1088/2041-8205/716/2/L109}, \href
  {http://adsabs.harvard.edu/abs/2010ApJ...716L.109M} {716, L109}

\bibitem[\protect\citeauthoryear{{Miller-Jones} et~al.,}{{Miller-Jones}
  et~al.}{2012}]{MJ2012}
{Miller-Jones} J.~C.~A.,  et~al., 2012, \mn@doi [\mnras]
  {10.1111/j.1365-2966.2011.20326.x}, \href
  {http://adsabs.harvard.edu/abs/2012MNRAS.421..468M} {421, 468}

\bibitem[\protect\citeauthoryear{{Mirabel} \& {Rodr{\'{\i}}guez}}{{Mirabel} \&
  {Rodr{\'{\i}}guez}}{1994}]{Mirabel1994}
{Mirabel} I.~F.,  {Rodr{\'{\i}}guez} L.~F.,  1994, \mn@doi [\nat]
  {10.1038/371046a0}, \href {http://adsabs.harvard.edu/abs/1994Natur.371...46M}
  {371, 46}

\bibitem[\protect\citeauthoryear{{Moore}, {Rutledge}, {Fox}, {Guerriero},
  {Lewin}, {Fender}  \& {van Paradijs}}{{Moore} et~al.}{2000}]{Moore2000}
{Moore} C.~B.,  {Rutledge} R.~E.,  {Fox} D.~W.,  {Guerriero} R.~A.,  {Lewin}
  W.~H.~G.,  {Fender} R.,   {van Paradijs} J.,  2000, \mn@doi [\apj]
  {10.1086/308589}, \href {http://adsabs.harvard.edu/abs/2000ApJ...532.1181M}
  {532, 1181}

\bibitem[\protect\citeauthoryear{{Mu{\~n}oz-Darias}, {Fender}, {Motta}  \&
  {Belloni}}{{Mu{\~n}oz-Darias} et~al.}{2014}]{MD2014}
{Mu{\~n}oz-Darias} T.,  {Fender} R.~P.,  {Motta} S.~E.,   {Belloni} T.~M.,
  2014, \mn@doi [\mnras] {10.1093/mnras/stu1334}, \href
  {http://adsabs.harvard.edu/abs/2014MNRAS.443.3270M} {443, 3270}

\bibitem[\protect\citeauthoryear{{Narayan} \& {Yi}}{{Narayan} \&
  {Yi}}{1995}]{Narayan1995}
{Narayan} R.,  {Yi} I.,  1995, \mn@doi [\apj] {10.1086/176343}, \href
  {http://adsabs.harvard.edu/abs/1995ApJ...452..710N} {452, 710}

\bibitem[\protect\citeauthoryear{{Negoro} et~al.,}{{Negoro}
  et~al.}{2015}]{ATel_MAXI2015}
{Negoro} H.,  et~al., 2015, The Astronomer's Telegram, \href
  {http://adsabs.harvard.edu/abs/2015ATel.7008....1N} {7008}

\bibitem[\protect\citeauthoryear{{Papitto} et~al.,}{{Papitto}
  et~al.}{2013}]{Papitto2013}
{Papitto} A.,  et~al., 2013, \mn@doi [\nat] {10.1038/nature12470}, \href
  {http://adsabs.harvard.edu/abs/2013Natur.501..517P} {501, 517}

\bibitem[\protect\citeauthoryear{{Parikh} et~al.,}{{Parikh}
  et~al.}{2017a}]{Parikh2016}
{Parikh} A.~S.,  et~al., 2017a, \mn@doi [\mnras] {10.1093/mnras/stw3388}, \href
  {http://adsabs.harvard.edu/abs/2017MNRAS.466.4074P} {466, 4074}

\bibitem[\protect\citeauthoryear{{Parikh}, {Wijnands}, {Degenaar},
  {Altamirano}, {Patruno}, {Gusinskaia}  \& {Hessels}}{{Parikh}
  et~al.}{2017b}]{Parikh2017}
{Parikh} A.~S.,  {Wijnands} R.,  {Degenaar} N.,  {Altamirano} D.,  {Patruno}
  A.,  {Gusinskaia} N.~V.,   {Hessels} J.~W.~T.,  2017b, \mn@doi [\mnras]
  {10.1093/mnras/stx747}, \href
  {http://adsabs.harvard.edu/abs/2017MNRAS.468.3979P} {468, 3979}

\bibitem[\protect\citeauthoryear{{Patruno} \& {Watts}}{{Patruno} \&
  {Watts}}{2012}]{Patruno2012}
{Patruno} A.,  {Watts} A.~L.,  2012, preprint, \href
  {http://adsabs.harvard.edu/abs/2012arXiv1206.2727P} {} (\mn@eprint {arXiv}
  {1206.2727})

\bibitem[\protect\citeauthoryear{{Perley}, {Chandler}, {Butler}  \&
  {Wrobel}}{{Perley} et~al.}{2011}]{VLA2011}
{Perley} R.~A.,  {Chandler} C.~J.,  {Butler} B.~J.,   {Wrobel} J.~M.,  2011,
  \mn@doi [\apjl] {10.1088/2041-8205/739/1/L1}, \href
  {http://adsabs.harvard.edu/abs/2011ApJ...739L...1P} {739, L1}

\bibitem[\protect\citeauthoryear{{Plotkin}, {Markoff}, {Kelly}, {K{\"o}rding}
  \& {Anderson}}{{Plotkin} et~al.}{2012}]{Plotkin2012}
{Plotkin} R.~M.,  {Markoff} S.,  {Kelly} B.~C.,  {K{\"o}rding} E.,   {Anderson}
  S.~F.,  2012, \mn@doi [\mnras] {10.1111/j.1365-2966.2011.19689.x}, \href
  {http://adsabs.harvard.edu/abs/2012MNRAS.419..267P} {419, 267}

\bibitem[\protect\citeauthoryear{{Rich}, {de Blok}, {Cornwell}, {Brinks},
  {Walter}, {Bagetakos}  \& {Kennicutt}}{{Rich} et~al.}{2008}]{MSCLEAN2008}
{Rich} J.~W.,  {de Blok} W.~J.~G.,  {Cornwell} T.~J.,  {Brinks} E.,  {Walter}
  F.,  {Bagetakos} I.,   {Kennicutt} Jr. R.~C.,  2008, \mn@doi [\aj]
  {10.1088/0004-6256/136/6/2897}, \href
  {http://adsabs.harvard.edu/abs/2008AJ....136.2897R} {136, 2897}

\bibitem[\protect\citeauthoryear{{Romano} et~al.,}{{Romano}
  et~al.}{2006}]{ROM2006}
{Romano} P.,  et~al., 2006, \mn@doi [Nuovo Cimento B Serie]
  {10.1393/ncb/i2007-10062-y}, \href
  {http://adsabs.harvard.edu/abs/2006NCimB.121.1067R} {121, 1067}

\bibitem[\protect\citeauthoryear{{Rupen}, {Dhawan}, {Mioduszewski}, {Stappers}
  \& {Gaensler}}{{Rupen} et~al.}{2002}]{Rupen2002}
{Rupen} M.~P.,  {Dhawan} V.,  {Mioduszewski} A.~J.,  {Stappers} B.~W.,
  {Gaensler} B.~M.,  2002, \iaucirc, \href
  {http://adsabs.harvard.edu/abs/2002IAUC.7997....2R} {7997}

\bibitem[\protect\citeauthoryear{{Rushton} et~al.,}{{Rushton}
  et~al.}{2016}]{Rushton2016}
{Rushton} A.~P.,  et~al., 2016, \mn@doi [\mnras] {10.1093/mnras/stw2020}, \href
  {http://adsabs.harvard.edu/abs/2016MNRAS.463..628R} {463, 628}

\bibitem[\protect\citeauthoryear{{Russell}, {Miller-Jones}, {Maccarone},
  {Yang}, {Fender}  \& {Lewis}}{{Russell} et~al.}{2011}]{Russell2011}
{Russell} D.~M.,  {Miller-Jones} J.~C.~A.,  {Maccarone} T.~J.,  {Yang} Y.~J.,
  {Fender} R.~P.,   {Lewis} F.,  2011, \mn@doi [\apjl]
  {10.1088/2041-8205/739/1/L19}, \href
  {http://adsabs.harvard.edu/abs/2011ApJ...739L..19R} {739, L19}

\bibitem[\protect\citeauthoryear{{Rutledge}, {Moore}, {Fox}, {Lewin}  \& {van
  Paradijs}}{{Rutledge} et~al.}{1998}]{Rutledge1998}
{Rutledge} R.,  {Moore} C.,  {Fox} D.,  {Lewin} W.,   {van Paradijs} J.,  1998,
  \iaucirc, \href {http://adsabs.harvard.edu/abs/1998IAUC.6813....2R} {6813}

\bibitem[\protect\citeauthoryear{{Shakura} \& {Sunyaev}}{{Shakura} \&
  {Sunyaev}}{1973}]{SS1973}
{Shakura} N.~I.,  {Sunyaev} R.~A.,  1973, \aap, \href
  {http://adsabs.harvard.edu/abs/1973A%26A....24..337S} {24, 337}

\bibitem[\protect\citeauthoryear{{Stappers} et~al.,}{{Stappers}
  et~al.}{2014}]{Stappers2014}
{Stappers} B.~W.,  et~al., 2014, \mn@doi [\apj] {10.1088/0004-637X/790/1/39},
  \href {http://adsabs.harvard.edu/abs/2014ApJ...790...39S} {790, 39}

\bibitem[\protect\citeauthoryear{{Tendulkar} et~al.,}{{Tendulkar}
  et~al.}{2014}]{Tendulkar2014}
{Tendulkar} S.~P.,  et~al., 2014, \mn@doi [\apj] {10.1088/0004-637X/791/2/77},
  \href {http://adsabs.harvard.edu/abs/2014ApJ...791...77T} {791, 77}

\bibitem[\protect\citeauthoryear{{Tetarenko} et~al.,}{{Tetarenko}
  et~al.}{2016}]{Tetarenko2016}
{Tetarenko} A.~J.,  et~al., 2016, \mn@doi [\mnras] {10.1093/mnras/stw1013},
  \href {http://adsabs.harvard.edu/abs/2016MNRAS.460..345T} {460, 345}

\bibitem[\protect\citeauthoryear{{Tudor} et~al.,}{{Tudor}
  et~al.}{2017}]{Tudor2017}
{Tudor} V.,  et~al., 2017, preprint, \href
  {http://adsabs.harvard.edu/abs/2017arXiv170505071T} {} (\mn@eprint {arXiv}
  {1705.05071})

\bibitem[\protect\citeauthoryear{{Tudose}, {Fender}, {Linares}, {Maitra}  \&
  {van der Klis}}{{Tudose} et~al.}{2009}]{TUD2009}
{Tudose} V.,  {Fender} R.~P.,  {Linares} M.,  {Maitra} D.,   {van der Klis} M.,
   2009, \mn@doi [\mnras] {10.1111/j.1365-2966.2009.15604.x}, \href
  {http://adsabs.harvard.edu/abs/2009MNRAS.400.2111T} {400, 2111}

\bibitem[\protect\citeauthoryear{{Wijnands} et~al.,}{{Wijnands}
  et~al.}{2006}]{Wijnands2006}
{Wijnands} R.,  et~al., 2006, \mn@doi [\aap] {10.1051/0004-6361:20054129},
  \href {http://adsabs.harvard.edu/abs/2006A%26A...449.1117W} {449, 1117}

\makeatother
\end{thebibliography}

\bsp	
\label{lastpage}
\end{document}